\documentclass[prd,a4paper,superscriptaddress,aps,showpacs,floats,nofootinbib,linenumbers]{revtex4-1}
\usepackage{epsfig}
\usepackage{graphicx}
\usepackage{amsmath}
\usepackage{amssymb}
\usepackage{color}
\usepackage{hyperref}
\usepackage{url}
\usepackage{float}
\usepackage{placeins}

\let\Oldsection\section
\renewcommand{\section}{\FloatBarrier\Oldsection}

\let\Oldsubsection\subsection
\renewcommand{\subsection}{\FloatBarrier\Oldsubsection}

\let\Oldsubsubsection\subsubsection
\renewcommand{\subsubsection}{\FloatBarrier\Oldsubsubsection}
\usepackage{subcaption}
\usepackage{setspace}

\makeatletter
\gdef\@ptsize{2} 
\makeatother

\begin{document}

\nolinenumbers
\title{Evaluating the Effect of Four Unknown Parameters Included in a Latitudinal Energy Balance Model on the Habitability of Exoplanets}

\author{M. Bahraminasr}
\email{m.bahrami.nasr@iota-me.com}

\affiliation{The International Occultation Timing Association-Middle East Section (IOTA/ME), Iran}

\author{S.J. Jafarzadeh}
\email{j.jafarzadeh@iota-me.com}

\affiliation{The International Occultation Timing Association-Middle East Section (IOTA/ME), Iran}

\author{F. Montazeri}
\email{f.montazeri.n@iota-me.com}

\affiliation{The International Occultation Timing Association-Middle East Section (IOTA/ME), Iran}

\author{A. Poro}
\email{info@iota-me.com}

\affiliation{The International Occultation Timing Association-Middle East Section (IOTA/ME), Iran}

\author{S. Sarabi}
\email{s.sarabi@iota-me.com}

\affiliation{The International Occultation Timing Association-Middle East Section (IOTA/ME), Iran}


\begin{doublespace}
\begin{abstract}

 Among different models for determining the habitable zone (HZ) around a star, a Latitudinal Energy Balance Model (LEBM) is very beneficial due to its parametricity which keeps a good balance between complexity and simulation time. This flexibility makes the LEBM an excellent tool to assess the impact of some key physical parameters on the temperature and the habitability of a planet. Among different physical parameters, some of them, up until now, cannot be determined by any method such as the planet's spin obliquity, diurnal period, ocean-land ratio, and pressure level. Here we apply this model to study the effect of these unknown parameters on the habitability of three exoplanets located in the inner, outer, and middle of their optimistic HZ. Among the examined parameters, the impact of pressure is more straightforward. It has a nearly direct relation with temperature and also with the habitability in the case of a cold planet. The effect of other parameters is discussed with details. To quantify the impact of all these unknown parameters we utilize a statistical interface which provides us with the conditional probability on habitability status of each planet.

\end{abstract}

\maketitle

\date{\today}


\section{Introduction}
\label{intro}

The first confirmation of an exoplanet orbiting a main-sequence star was made in 1995 when a giant planet was found in a four-day orbit around the nearby star 51 Pegasi. It was named 51 Pegasi b (Mayor et al., 1995). Following the discovery of 51 Pegasi b the field of exoplanet detection flourished and resulted in launching many projects, especially those based on transits. By now, over 2000 exoplanets have been detected by the Kepler telescope, and the numbers will expand further as a result of other projects like TESS (Transiting Exoplanet Survey Satellite) (Fischer et al., 2015).

   Due to the crucial role of liquid water in the biochemistry of life on the Earth, a habitable zone is defined as the spatial extent around a main-sequence star in which the planet can maintain liquid water on its surface over an extended period of time (Kasting et al., 1993). There are different models of studying the habitability of exoplanets, each one with different assumptions. Based on the complexity and dimensions of a model, it contains its own unique set of parameters. Kasting et al. (1993) calculated HZ limits for stars with effective temperatures higher than 3700 K employing a 1-D radiative-convective, cloud-free climate model. Kopparapu et al. (2013) updated HZ limits for stars with effective temperatures between 2600 K and 7200 K to estimate the actual occurrence rate of the Earth-size planets in the HZs around M-dwarfs. Moreover, Kopparapu et al. (2014) have constructed both conservative and optimistic estimates of the HZ from 1D and 3D models.
   
   Heng et al. (2015) focused on atmospheric quantities of highly irradiated large exoplanets or Hot Jupiters. They concluded that the most important implication for the habitability of these planets is the equator-to-pole temperature difference response to the variation in orbital and atmospheric parameters. Kopparapu et al. (2016) estimated the inner edge of the HZ for synchronously rotating terrestrial planets around late-K and M-dwarf stars with self-consistent relationships between stellar metallicity, stellar effective temperature, and the planetary orbital and rotational period. Furthermore, Kopparapu et al. (2017) studied ocean-covered planets at the inner edge of the HZ around late M to mid-K stars with a 3D climate system model. Yang et al. (2013) showed that the atmospheric circulation state affects horizontal heat transport and spatial distribution of clouds which both have an impact on habitability. In addition, planet rotation speed, and ,consequently, the Coriolis force have a significant effect on the planetary albedo and thus the climate of the exoplanet (Yang et al., 2014).
   
   In addition to the Kasting definition of a spatial HZ, there is another way of assessing the concept by evolving a climate model for large numbers of planets on a multidimensional grid of orbital elements, mapping out an ``orbital HZ'' in this parameter space (Forgan., 2013). An Energy Balance Model (EBM) is one of the above-mentioned climate models. After Budyko et al., (1969) and Sellers et al., (1969) a LEBM proved to be useful in climate science. After Williams et al., (1997), Franck et al., (2000), and Gaidos et al., (2004) this model has been utilized in terrestrial exoplanets. Moreover, in Spiegel et al., (2008) Sellers, W. D., (2011), and Vladilo et al., (2013) it was used for an Earth-like planet with different rotational speeds and various pressures, respectively.
   
   A variety of other works has studied the dependence of climate on surface pressure in 1D or 3D models. Goldblatt et al. (2009) studied the effect of doubling the present atmospheric nitrogen level using a radiative–convective model. Kopparapu et al. (2014) showed that larger planets have wider HZs. Keles et al. (2018) demonstrated that increasing pressure would result in increasing the temperature due to the greenhouse effect. However, in around the pressure of 4 bar, Rayleigh scattering gradually becomes dominated and thus temperature decreases. Charnay et al. (2013) used a 3D global climate model (GCM) to investigate the effect of atmospheric pressure as different solutions to the faint young Sun problem over time. Wolf \& Toon (2014) investigated the habitability of the Earth considering the increment of the Sun’s brightness. Chemke et al. (2016) showed that higher pressures tend to increase the near-surface temperature. Rushby et al. (2019) determined the effect of varying fractional and latitudinal distribution of land as a function of host star spectral energy distribution on the overall planetary albedo, climate, and ice-albedo feedback response. Colose et al (2019) showed that increasing obliquity results in increasing the temperature of the planet because of ice-albedo feedbacks in cold climates and water vapor warm climates. 

   Some of the physical orbit parameters like semi-major axis and eccentricity, which affect habitability, can be achieved by observation. However, many parameters are still unknown. Among them are atmospheric pressure, diurnal period, spin obliquity, and ocean coverage.

   For an orbital HZ, we are free to manipulate some parameters which makes this a suitable tool to speculate on cases having different ocean coverages or different spin obliquities. It may create an issue on how changing unknown parameters can change the habitable fraction of a planet and whether this can change the status of a planet from habitable to unhabitable or vice versa.
   
   This paper is organized as follows: Section \ref{lebm} introduces the LEBM, section \ref{excasestudy} presents our exoplanet case studies, section \ref{discussion} discusses the effect of each parameter on the temperature and habitability, and section \ref{conclusion} concludes our results.


\nolinenumbers
\section{Latitudinal Energy Balance Model}
\label{lebm}

   The simplest method of climate models, the zero-dimensional model, considers the planet as a particle satisfying energy conservation which could be used to solve the planet’s effective temperature. This temperature could be different from the surface temperature based on the greenhouse effect of the atmospheric gases that absorb thermal radiation. To have a more realistic view of the climate, the temperature should be latitudinally resolved as a one-dimensional climate model. A LEBM or zonally averaged EBM treats each zonal strip separately and, thus, the rate of transporting of energy between different layers is also considered. 
   
   Following previous works, Williams et al., (1997) and Spiegel et al., (2008) the main equation of a LEBM is a one-dimensional heat diffusion equation which relates incoming, reflecting and outgoing radiation to the temperature of each layer.

\begin{equation}
\begin{split}
C   \frac{\partial T[x,t]}{\partial t}&-\frac{\partial}{\partial{x}}\Big(D(1-x^2) \frac{\partial T[x,t]}{\partial x}\Big) 
 = S(1-A[T])-I[T]
\end{split}
\label{r1}
\end{equation}

$C$ is the effective heat capacity, $T$ is temperature, $D$ is the diffusion coefficient, which determines the efficacy of zonal exchange heat, $x$ is related to the latitude such that $\lambda = sin^{-1}x$ , $I$ is Outgoing Long wave Radiation (OLR),  $S$ is instellation, and $A$ represents the albedo of the layer. The $(1-x^2)$ term arises from solving diffusion equation in spherical geometry.

\nolinenumbers

\subsection{Model Parameters}

   To Include the pressure, we follow the definitions of Vladilo et al., (2013) for heat capacity, diffusion coefficient, albedo, and OLR, which are as follows.
 

\nolinenumbers
\subsubsection{Heat Capacity}

   The atmosphere thermal inertia depends on the fraction of the planet's surface which is covered with ocean, land, and ice. Considering different levels of pressure, the heat capacity of the atmosphere becomes:
 \begin{equation}
 C_{atm}=\Big( \frac{c_{p}}{c_{p,o}}\Big)\Big( \frac{P}{P_{o}}\Big)C_{atm,o}
 \label{r2}
 \end{equation}
where $c_p$ and $P$ are specific heat capacity and pressure of the atmosphere. Index $o$ in this relation corresponds to the values for the Earth. Heat capacity of the atmosphere, $C_{atm,o}$ is set to be $10.1\times 10^{6}  \text{J} \text{ m}^{-2}\text{ K}^{-1}$, and heat capacities of land, $C_l$, ocean, $C_o$, and ice, $C_i$ are defined as follows:
\begin{equation}
\begin{split}
C_l&=10^6+C_{atm,o}\\
C_o&=210 \times10^6+C_{atm,o}\\
C_{i}&=
  \begin{cases}
    1.0\times10^6 + C_{atm} & T<263 \\
    43\times10^6 + C_{atm} & 263<T<273
  \end{cases}
\end{split}
 \label{r3}
\end{equation}
The total heat capacity is equal to
\begin{equation}
C=f_{l}C_l+f_o[(1-f_{i}) C_o + f_i C_i]
 \label{r4}
\end{equation}
The relation between $f_l$, fraction of land, and $f_o$, fraction of ocean, is
\begin{equation}
f_l = 1 - f_o
 \label{r5}
\end{equation}
The fraction of the surface covered by ice is determined by
\begin{equation}
f_i(T)=max\Big\{0,\big[1-e^{\frac{T-273}{10 K}}\big]\Big\}
 \label{r6}
\end{equation}


\nolinenumbers
\subsubsection{Diffusion Coefficient}

 The diffusion coefficient $D$ which approximate the heat transport by atmospheric circulation is defined such that a planet at 1 A.U. around a star of mass $1\text{M}_{\odot}$ with a rotational period of 1 day will reproduce the average temperature profile measured on the Earth. For a planet with a different rotational period, $D$ is proportional to the rotational velocity ${\omega_o}^{-2}$ and for different pressures, $D$  is linearly proportional to the pressure. The diffusion coefficient is thus:
\begin{equation}
D=\Big(\frac{P}{P_o}\Big)\Big(\frac{c_P}{c_{P,o}}\Big)\Big(\frac{m}{m_o}\Big)^{-2}\Big(\frac{\omega}{\omega_o}\Big)^{-2} D_o
 \label{r7}
\end{equation}
$\omega$ is the rotational velocity of the planet, and $m$ is the mean molecular weight. Index $o$ corresponds to the values of the Earth (Williams \& Kasting, 1997).


\nolinenumbers
\subsubsection{Albedo}
\label{Albedo}

   Albedo is defined based on the fraction of land, $f_{l}$, ocean, $f_{o}$, ice on them, $f_{i}$, the fraction of clouds on land, $f_{cl}$, clouds on water, $f_{cw}$, and clouds on ice, $f_{ci}$.
\begin{equation}
\begin{split}
A= & f_o \bigg \{(1-f_i)\Big[a_o(1-f_{cw}) +  a_cf_{cw}\Big] 
+ f_i\Big[a_{io}(1-f_{ci})+a_c f_{ci}\Big]\bigg \} \\
+& f_l\bigg \{(1-f_i)\Big[a_l(1-f_{cl})+a_c f_{cl}\Big]  
+ f_i\Big[a_{il}(1+f_{ci})+a_c f_{ci}\Big]\bigg \} 
\end{split}
 \label{r8}
\end{equation}
This equation includes parameters which are defined as:
\begin{equation}
\begin{split}
&a_o = \frac{0.026}{(1.1\mu^{1.7}+0.065)}+0.15(\mu-0.1)(\mu-0.5)(\mu-1.0) \\
&\mu = \cos Z_{*} \\
& a_c = \text{max}\Big\{a_{c0},[\alpha+\beta Z_*]\Big\} \\
& \alpha = -0.07, \text{\vspace{1cm} }\beta=8\times10^{-3} (^{\circ})^{-1} \\
& a_{il} = 0.85,\text{\vspace{1cm} }  a_{io}=0.62, \text{\vspace{1cm} } a_l = 0.2 \\
& f_{cw}=0.67, \text{\vspace{1cm} } f_{cl}=0.50, \text{\vspace{1cm} } f_{ci}=0.50
\end{split}
 \label{r9}
\end{equation}

$a_{o}$, $a_{l}$, and $a_{c}$ are the albedos of ocean, land, and clouds respectively. $a_{io}$ and $a_{il}$ are the albedos of the ice on the ocean and ice on the land. $Z_{*}$ is the zenith distance of the star and $\alpha$ and $\beta$ are estimated such that the relation $\alpha+\beta Z_*$ yield a good fit to Cess data. (Cess, R.D., 1976). For a more detailed description of each parameter see Vladilo et al., (2013).

  Implementing this relation for albedo instead of the temperature dependent relation in Spiegel et al., (2009) prevents the Earth from freezing when the simulation starts at the northern winter solstice or when the solar distance is changed from 1 AU to 1.025 AU. In each configuration, the Earth freezes as a consequence of choosing an approximation for albedo which fails to predict some situations correctly (Appendix \ref{AppNewAlbido}).


\nolinenumbers
\subsubsection{Outgoing Long wave Radiation}

The OLR is defined as:
\begin{equation}
I=\frac{\sigma T^4}{1+0.75\tau_{IR}[T,P]}
 \label{r10}
\end{equation}
Where $\sigma$ is the Boltzmann constant and $\tau_{IR}$ is the optical depth of the atmosphere which for atmospheric pressure is defined as:
\begin{equation}
\tau_{IR}(T, P = 1\ atm) = 0.79 \Big(\frac{T}{273 K}\Big)^3
 \label{r11}
\end{equation}
   Changing the pressure represents itself in the optical depth which governs the greenhouse effect. It is expected that $\tau_{IR}$ has a strong dependence on P since $\tau_{IR}=\kappa_{IR}\ p / g$ where $\kappa_{IR}$ is the total absorption coefficient  and g the gravitational acceleration. Furthermore, in the range of planetary surface pressures, $\kappa_{\mathrm{IR}}$ is dominated by collisional broadening, which introduces a linear dependence on $P$ of the widths of the absorption lines (Pierrehumbert, 2010; Salby 2012). In absence of line saturation, $\kappa_{\mathrm{IR}}$ should therefore increase linearly with $P$. Therefore, one might expect that $\tau_{\mathrm{IR}}\propto P^2$; however, it is milder in reality and deriving an analytical function is not possible. It is required to perform a series of radiative calculations and tabulate the results as a function of T and P. The detailed description of the procedure can be found in Vladilo et al., (2013).


\nolinenumbers
\subsubsection{Incoming Solar Radiation}

   Assuming a simple main sequence scaling for the luminosity, the total averaged diurnal instellation becomes:
\begin{equation}
S = q_o \Big( \frac{H}{\pi}\Big) \bar{\mu} = \frac{q_o}{\pi}(H \sin \lambda \sin \delta + \cos \lambda \cos \delta \sin H )
 \label{extra_r100}
\end{equation}
$\bar{\mu}$ is the mean diurnal value of $\mu = \cos Z_*$ when the star is above the horizon. The bolometric flux received at a distance of 1 A.U. becomes:
\begin{equation}
 q_o = 1.36 \times 10^6 \Big( \frac{M_{*}}{M_{\odot}}\Big)^4 erg s^{-1} cm^{-2}
 \label{extra_r101}
\end{equation}
The radian half-day length (H) is:
\begin{equation}
\cos H = - \tan \lambda \tan \delta
 \label{extra_r102}
\end{equation}
$\delta$ is the solar declination which is calculated as follows:
\begin{equation}
\sin \delta = - \sin \delta_o \cos(\phi_p -\phi_{peri} - \phi_a)
 \label{extra_r103}
 \end{equation}

$\delta_o$ is the obliquity, $\phi_p$ is the current orbital longitude of the planet, $\phi_{peri}$ is the longitude of periastron, and $\phi_a$ is the longitude of winter solstice relative to the longitude of periastron.

\nolinenumbers


\subsection{Simulation Procedure and the Habitability}

In order to solve equation  \ref{r1} we used a staggered grid in which variables are calculated at the center of the grid cells and their derivatives at the cells borders. The spatial resolution is $1.25^{\circ}$ (145 grid points from the north to the south pole), and a global time step was adopted with a stable time step constraint for equation \ref{r1}:
\begin{equation}
\delta t < \frac{(\Delta x)^2 C}{2D(1-x^2)}
 \label{r15}
\end{equation}
The boundary condition is $\frac{\mathrm{d}T}{\mathrm{d}\lambda}=0 $ at the poles.

The habitability function of each latitude is defined as:
\begin{equation}
\begin{split}
h[\lambda, t]=
\begin{cases}
1&273 \leq T[\lambda,t] \leq T_{boil}(p_o)\\
0  &\text{otherwise}
\end{cases}
\end{split}
 \label{r13}
\end{equation}
The fraction of the potentially habitable surface at time $t$ is:
\begin{equation}
H_{area}[\lambda]=\frac{\int^{\pi/2}_{-\pi/2}h[\lambda, t]\cos \lambda \mathrm{d} \lambda}{2}
 \label{r14}
\end{equation}

To define habitability status, we use equation \ref{r13} which calculates the habitable fraction of the surface. When a steady-state is reached, we calculate mean $\bar{H}$ and the standard deviation $\sigma_{H}$ of the habitability fraction. Finally, the habitability status of the planet is determined according to Table \ref{t1} in which we classify the status of the planet as habitable, hot, snowball, or transient (Forgan., 2013).

\begin{table}
\begin{center}
  \caption{Habitability status}
  \begin{tabular}{ | l | c |  }
    \hline
    Condition & State  \\ \hline
    $\bar{H}>0.1\ \text{and} \ \sigma_{H} < 0.1\bar{H}$ & Habitable Planet \\ \hline
      $\bar{H}<0.1\ \text{and} \ T>373\text{K for all seasons}$ & Hot Planet \\ \hline
     $\bar{H}<0.1\ \text{and} \ \text{completely frozen}$ & Snowball Planet \\ \hline
     $\bar{H}>0.1\ \text{and} \ \sigma_{H} > 0.1\bar{H}$& Transient Planet \\
    \hline
  \end{tabular}
  \label{t1}
\end{center}
\end{table}

\nolinenumbers


\section{exoplanet case study}
\label{excasestudy}

Our goal is to study the impact of different unknown parameters of an exoplanet on its temperature and habitability. These parameters include pressure, the fraction of ocean and land, obliquity, and diurnal period. Usually, these parameters are set to the values of the Earth parameters. However, it is just one case among numerous cases which have different values. We want to study as many cases as possible in order to see how planet habitability behaves with a change in those parameters.

   To show this effect, we choose three planets --- $\tau$ Ceti e, Kepler-22b, and Kepler-62f --- among the discovered exoplanets. These three planets according to Kopparapu et al., (2013) and (2014) are respectively on the inner, middle, and outer edge of the optimistic habitable zone. Therefore, each one serves as a representation of other exoplanets located in the same approximate location in their star HZ.

The planet $\tau$  Ceti e is on the inner edge of the optimistic HZ of a star with a mass of 0.783$\pm$0.012 $\text{M}_{\odot}$, and a temperature T=5344 $\pm$ 50 K. The planet has a semi-major axis of 0.552 AU, eccentricity of 0.05, and orbital period of 162.87 days (Teixeira, T.C., et al., 2009). The minimum mass of this planet is 3.94 $M_{\oplus}$ or 0.0124 $M_{j}$ ($\oplus$ and j refer to the Earth and Jupiter respectively) and its radius is not defined (Feng et al., 2017). The planet Kepler-22b, designated as KOI-087.01, (Borucki, W. J., 2011) orbits within the optimistic HZ of the Sun-like star Kepler-22 with mass 0.970 M$_{\odot}$ and effective temperature $T_{eff} =5581 \pm 44 K$. The planet is located in an orbit with a semi-major axis of 0.849 AU, zero eccentricity, and orbital period of 286.89 days. This planet has a maximum mass of 35.87 $M_{\oplus}$ (0.113 $M_{j}$) and radius of 2.38 $R_{\oplus}$ (0.212 $R_{j}$) (Borucki et al., 2012 and Torres, G., et al., 2015).

 The planet Kepler-62f , designated as KOI-701, (Brown, T.M., et al., 2011) is located on the outer edge of the optimistic HZ of a star with mass 0.69 $M_{\odot}$ and effective temperature $T_{eff} = 4925 \pm 70 K$. The planet has a semi-major axis of 0.718 AU, zero eccentricity and orbital period of 267 days. Kepler-62f has a maximum mass of 34.92 $M_{\oplus}$ (0.110 $M_{j}$) and radius of 1.41 $R_{\oplus}$ (0.126 $R_{j}$) (Borucki et al., 2013), (Akeson, R. L., et al., 2013). Kepler-62f is probably a rocky world because of its Earth-like radius . The radius of Kepler-22b is in the range of a gas dwarf or mini-Neptune, however, being a rocky world without a very dense atmosphere is also possible (Buchhave, L.A. et al., 2014 and Kane, S.R., et a., 2016).


\nolinenumbers
\section{Results \& Discussion}
\label{discussion}

In the simulation, the spin obliquity (S.O.) is varied to (0.0, 22.5, 45.0, 67.5, 90.0) degrees, diurnal period (D.P.) to (0.5, 0.75, 1.0, 1.25, 1.5) days, ocean fraction (O.F.) to (0.1, 0.325, 0.55, 0.775, 1.0), pressure (P) to (0.1, 0.4, 0.7, 1.0, 1.3) atm for $\tau$ Ceti e and Kepler-22b, and (0.5, 1.5, 2.5, 3.5, 4.5) atm for Kepler-62f (The effect of changing these variables has been demonstrated in figures \ref{fig8} to \ref{fig11}). Since the model is longitudinally averaged, it can only be utilized in situations where the diurnal period is small compared to the orbital period. Considering this limitation, the range of diurnal period is relatively small to not to approach critical values. (Forgan., 2013). The reason for choosing different values of pressure for Kepler-62f was that in the range of other planets' pressures there was not a notable change in the temperature or the habitability of Kepler-62f. Then, raising the pressure to the aforementioned values showed its effect on temperature and habitability

   To present the effect of changing the parameters on the temperature and habitability, we monitored the changes at the end of the determined interval by fixing each parameter to its minimum and maximum and letting our parameter of interest to vary across the entire range of values.

   The effect of each parameter of interest on minimum, mean, and maximum temperatures and habitability is generally non-linear. Thus, the interpretation is not straight forward. However, an overall description tolerating a few deviations is possible. We explain this in the following sections.


\nolinenumbers
\subsection{Changing Pressure}

\begin{figure}[H]
    \centering
    \begin{subfigure}{0.49\textwidth}
        \includegraphics[width=\textwidth]{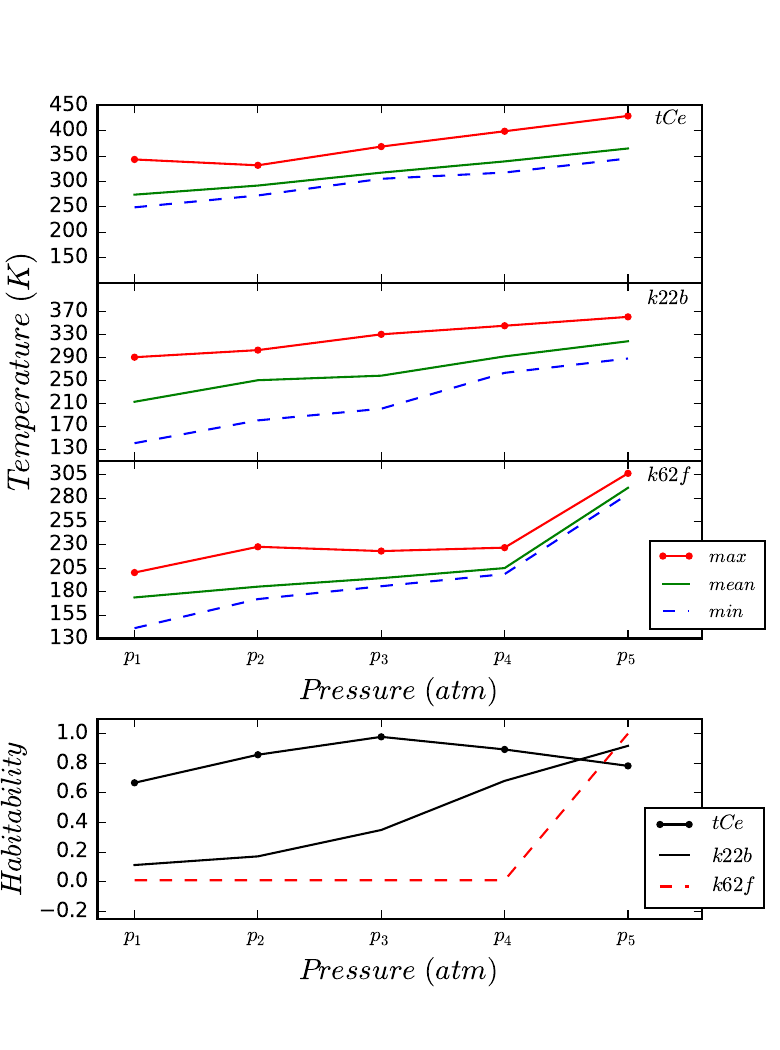}
        \caption{O.F. = 0.1, D.P. = 0.5 days, S.O. = 90.0 degrees}
        \label{fig8-a}
    \end{subfigure}
 \hfill 
    \begin{subfigure}{0.49\textwidth}
        \includegraphics[width=\textwidth]{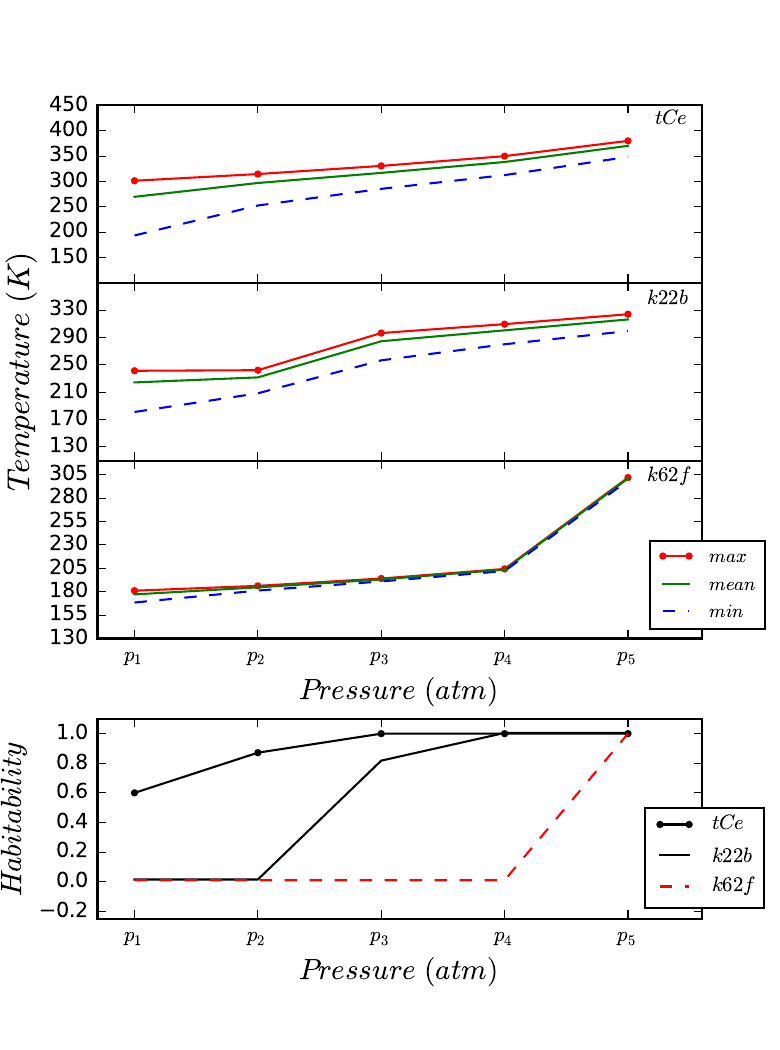}
        \caption{O.F. = 0.1, D.P. = 1.5 days, S.O. = 0.0 degrees}
        \label{fig8-b}
    \end{subfigure}
    \newline
 \hfill 
    \begin{subfigure}{0.49\textwidth}
        \includegraphics[width=\textwidth]{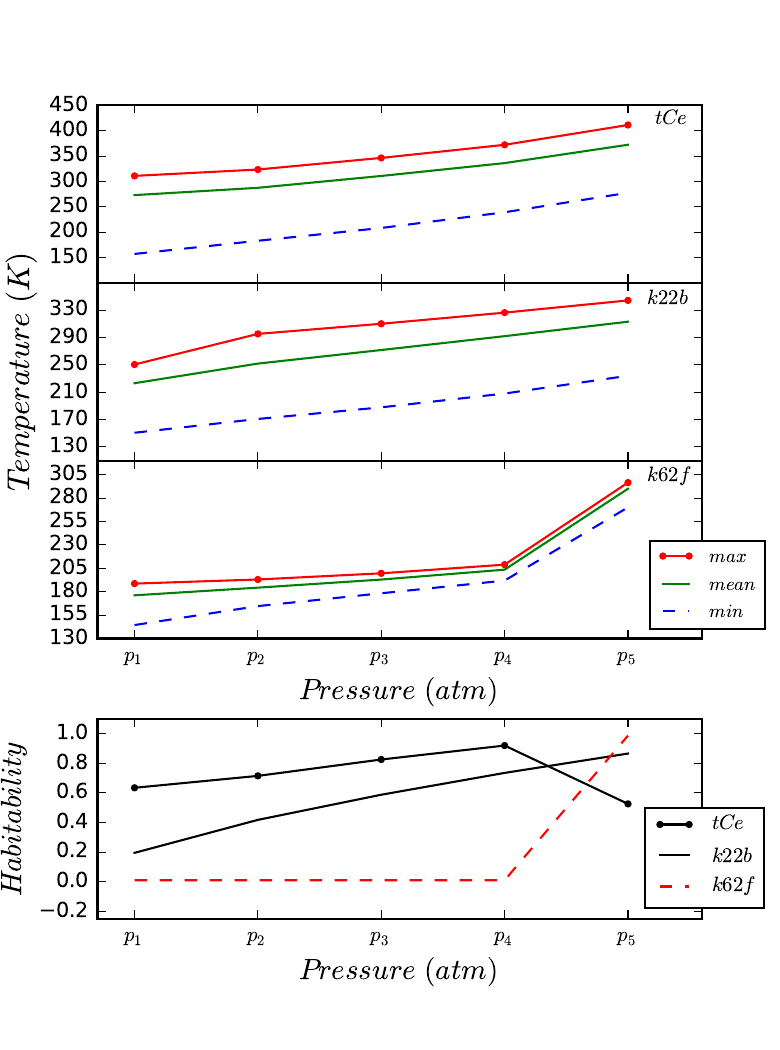}
        \caption{O.F. = 0.1, D.P. = 0.5 days, S.O. = 0.0 degrees}
        \label{fig8-c}
    \end{subfigure}
 \hfill 
    \begin{subfigure}{0.49\textwidth}
        \includegraphics[width=\textwidth]{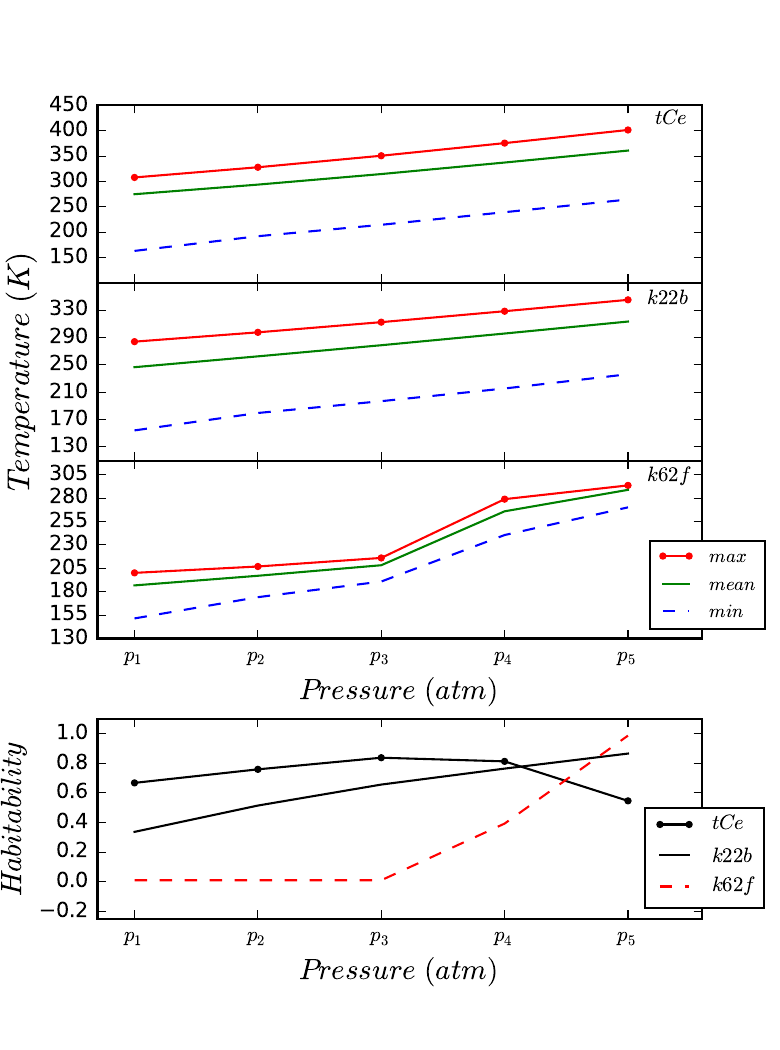}
        \caption{O.F. = 0.9, D.P. = 0.5 days, S.O. = 0.0 degrees}
        \label{fig8-d}
    \end{subfigure}
    \caption{Pressure varies; for planet $\tau$ Ceti e (tCe),  Kepler-22b (k22b), $p=(0.1, 0.4, 0.7,1.0, 1.3)$ atm; and for Kepler-62f (k62f), $ p=(0.5, 1.5,2.5, 3.5, 4.5)$ atm. }
    \label{fig8}
\end{figure}

   In LEBM, the main effect of increasing pressure is increasing the diffusion. Therefore, heat transfers between zonal strips more effectively. The result is that the process of glaciation would be reduced and it would lower the albedo which in turn increases the temperature. This effect is in accordance with the plots of figure \ref{fig8}. A planet like $\tau$ Ceti e is prone to be hot due to its location on the inner edge of the HZ. Therefore, increasing the pressure could decrease the habitability. In figure \ref{fig8-a}, the habitability fraction of this planet starts with a value of 0.67 at a pressure of 0.1 atm, increases to nearly one at a pressure of 0.7 atm, and decreases again to the value of 0.78 at a pressure of 1.3 atm. In the other two cases (Kepler-22b, Kepler-62f), what is observed is increasing the habitability, especially in the case of Kepler-62f, where the planet becomes unhabitable at low pressures. In figures \ref{fig8-a} \ref{fig8-b} \ref{fig8-c}, Kepler-62f has a zero habitability at pressures below 3.5 atm. However, when its pressure increases to 4.5 atm, its habitability has a significant increase and reaches the value of one.

   In more accurate models, various mechanism are included. In fact, The effect of surface temperature could not easily be estimated. Increasing surface pressure would result in the increment of albedo via Rayleigh scattering and also increases the greenhouse effect which implementing them is beyond the scope of this paper (Zsom, et al 2018). Furthermore, since increasing the pressure has a high impact on the temperature, other effects such as the moist and runaway greenhouse, which are not implemented in the standard LEBM, should be studied with more cautious. Those two effects are expected to occur on a planet with mean temperature above 350 K (Spiegel et al., 2008, Kopparapu et al., 2013).


\nolinenumbers
\subsection{Changing Ocean Fraction}

According to equation \ref{r4} , as the water has a higher amount of heat capacity in comparison to the land, increasing the ocean fraction will increase the total heat capacity. Regarding the required time for a planet to reach a semi steady-state temperature profile, the main effect of heat capacity is to slow down the changes in the temperature profile of the planet. However, if we allow enough time for the planet to evolve, it will finally reach its semi steady-state temperature profile regardless of the value of its heat capacity. This is evident in figure \ref{last}.

\begin{figure}[H]
    \centering
    \begin{subfigure}{0.49\textwidth}
        \includegraphics[width=\textwidth]{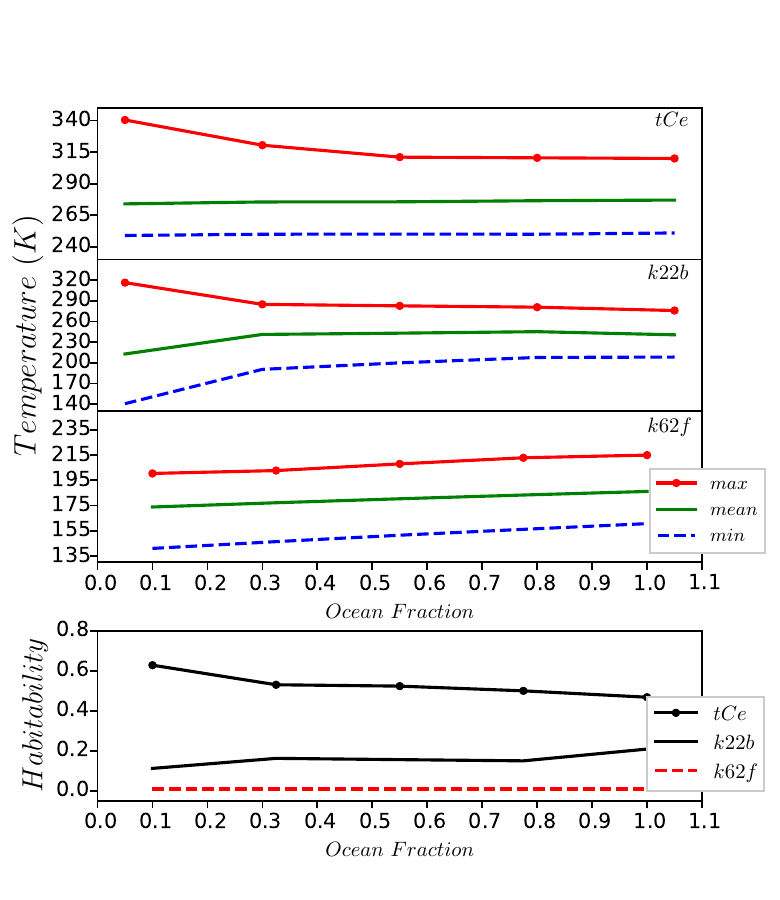}
        \caption{D.P. = 0.5 days, S.O. = 90.0 degrees, Pressure = $p_1$}
         \label{fig9-a}
    \end{subfigure}
 \hfill 
    \begin{subfigure}{0.49\textwidth}
        \includegraphics[width=\textwidth]{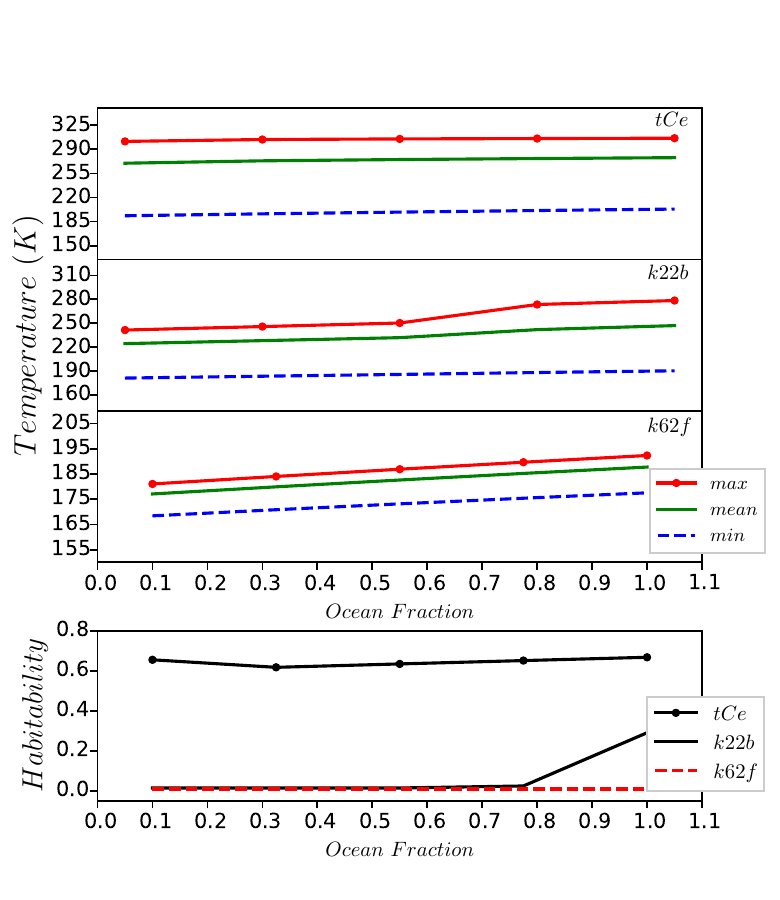}
        \caption{D.P. = 1.5 days, S.O. = 0.0 degrees, Pressure =  $p_1$}
        \label{fig9-b}
    \end{subfigure}
    \newline
 \hfill 
    \begin{subfigure}{0.49\textwidth}
        \includegraphics[width=\textwidth]{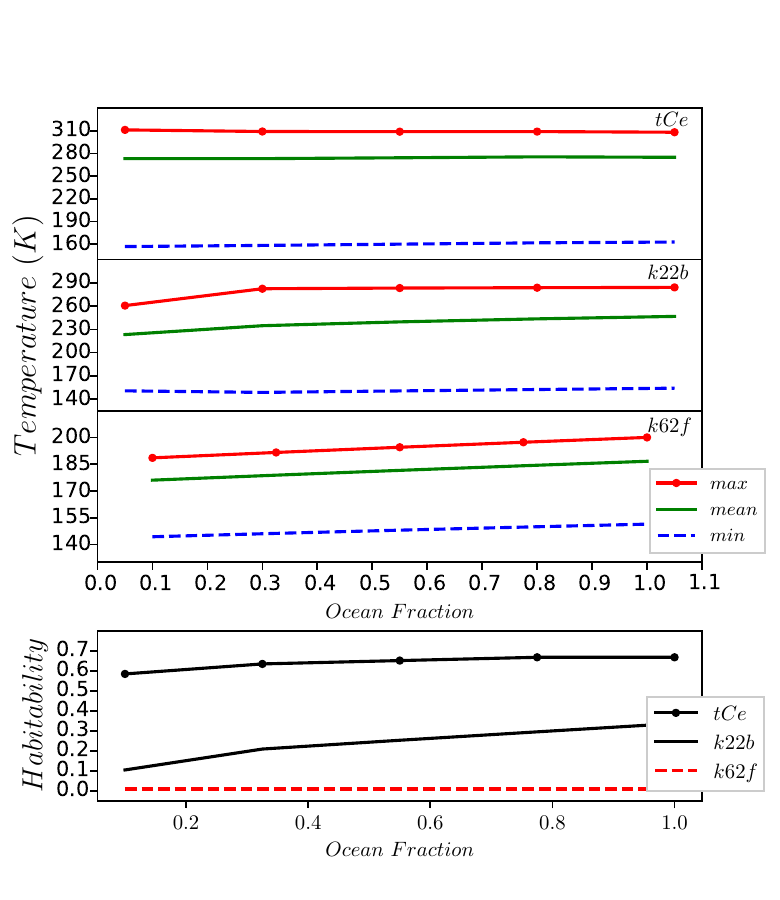}
        \caption{D.P. = 0.5 days, S.O. = 0.0 degrees, Pressure =  $p_1$}
        \label{fig9-c}
    \end{subfigure}
 \hfill 
    \begin{subfigure}{0.49\textwidth}
        \includegraphics[width=\textwidth]{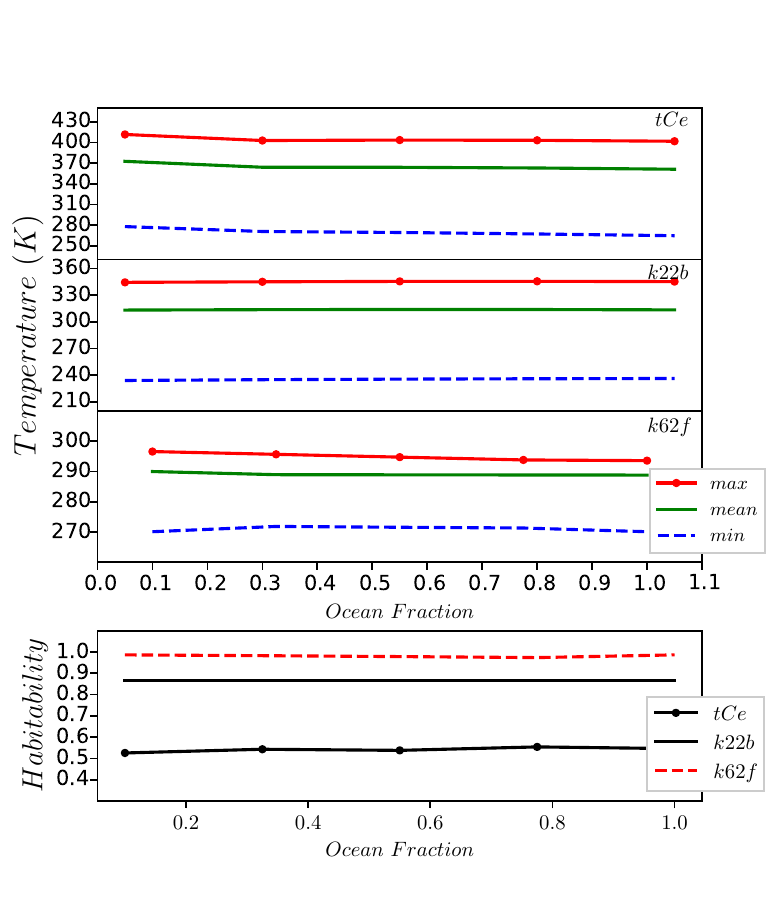}
        \caption{D.P. = 0.5 days, S.O. = 0.0 degrees, Pressure =  $p_5$}
        \label{fig9-d}
    \end{subfigure}
    \caption{Ocean fraction varies; for planet $\tau$ Ceti e (tCe),  Kepler-22b (k22b), $p_1 =0.1\ atm, \text{\ }p_5 =1.3\ atm$; and for Kepler-62f (k62f) $p_1\ =0.5 atm, \text{\ }p_5 =4.5\ atm$ }
    \label{fig9}
\end{figure}

  In low spin obliquities, increasing the ocean fraction generally results in slightly increasing the temperature due to the capability of maintaining more heat. This temperature increment makes planets more habitable in two ways. First, the habitability fraction increases for Kepler-22b, located in the middle region of the HZ (See figure  \ref{fig9} in which this increment ranges from 0.25 in state c to 0.3 in state b). Second, in addition to some increase in the habitability fraction of $\tau$ Ceti e, its habitability status changes from transient to habitable in some cases. Look at figure \ref{fig5} for these special cases.
  
   When the obliquity is very high, increasing the ocean fraction decreases the difference between maximum and minimum temperatures resulting in a flatter temperature profile. This narrowing of temperature differences is most evident for Kepler-22b in the case of figure  \ref{fig9-a} where the difference between these two temperatures changes from 176 to 68 degrees. In these cases, because of the particular configuration of the planet in its orbit, the ice belts created on the equator are weaker than ice caps in the case of zero obliquity. The reason is that while the star altitude at the poles of the planet in a zero obliquity case is always zero, it changes from 0-to-90 degrees at the equator of the planet with a 12-hour day cycle in the case of 90-degree obliquity based on the phase of the planet in its orbit. When the planet is semi-habitable (having a habitability fraction between zero and one), this ocean fraction increment leads to an increase in the habitability of the planet as in the case of Kepler-22b, where its habitability increases from 0.11 to 0.22 while it decreases for $\tau$ Ceti e as it is prone to get warmer.


\nolinenumbers
\subsection{Changing Spin Obliquity}
\label{spin_section}

\begin{figure}[H]
    \centering
    \begin{subfigure}{0.49\textwidth}
        \includegraphics[width=\textwidth]{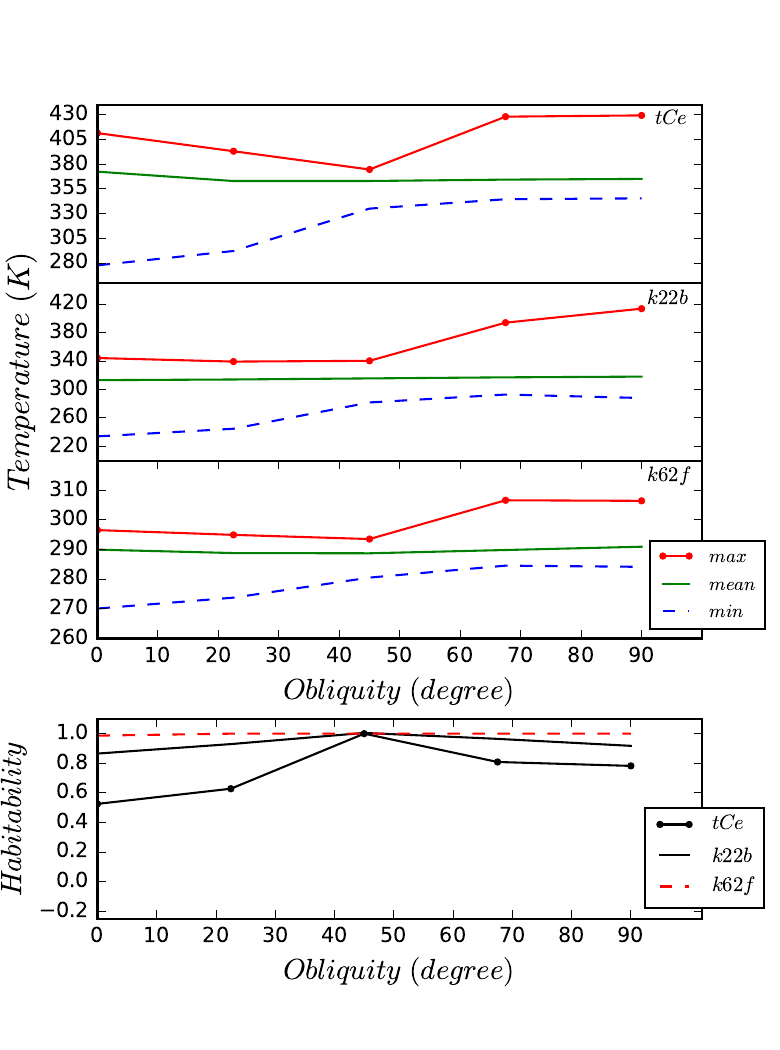}
        \caption{O.F. = 0.1, D.P. = 0.5 days, Pressure = $p_5$}
        \label{fig10-a}
    \end{subfigure}
 \hfill 
    \begin{subfigure}{0.49\textwidth}
        \includegraphics[width=\textwidth]{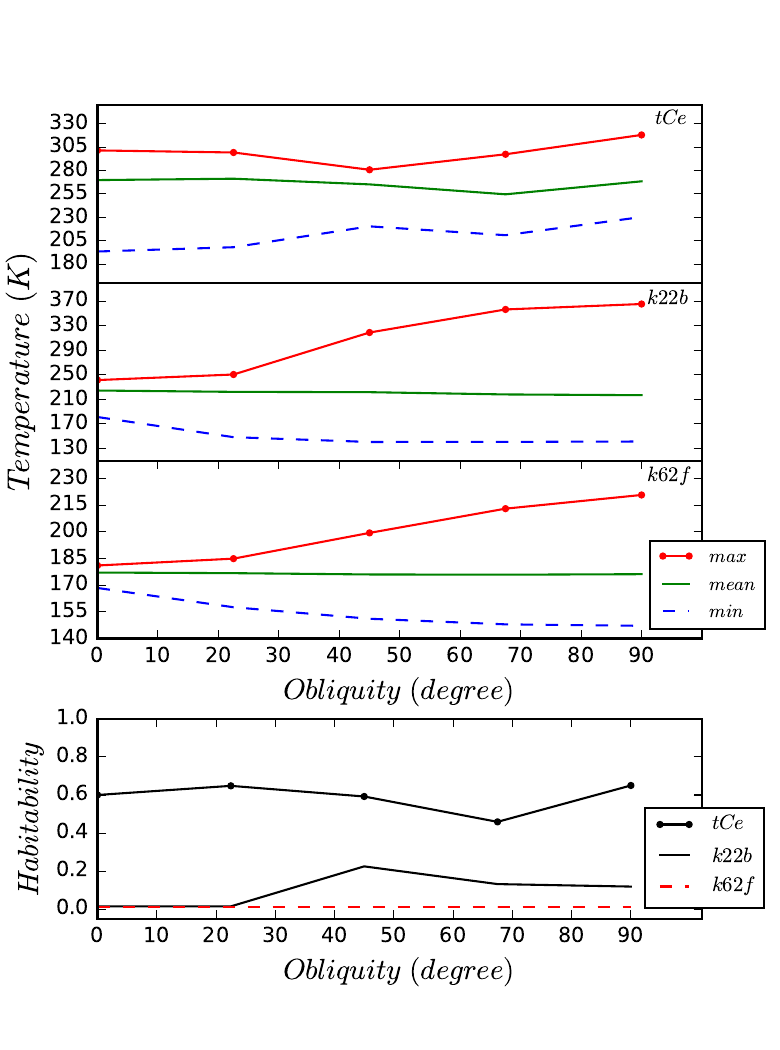}
        \caption{O.F. = 0.1, D.P. = 1.5 days, Pressure = $p_1$}
        \label{fig10-b}
    \end{subfigure}
    \newline
 \hfill 
    \begin{subfigure}{0.49\textwidth}
        \includegraphics[width=\textwidth]{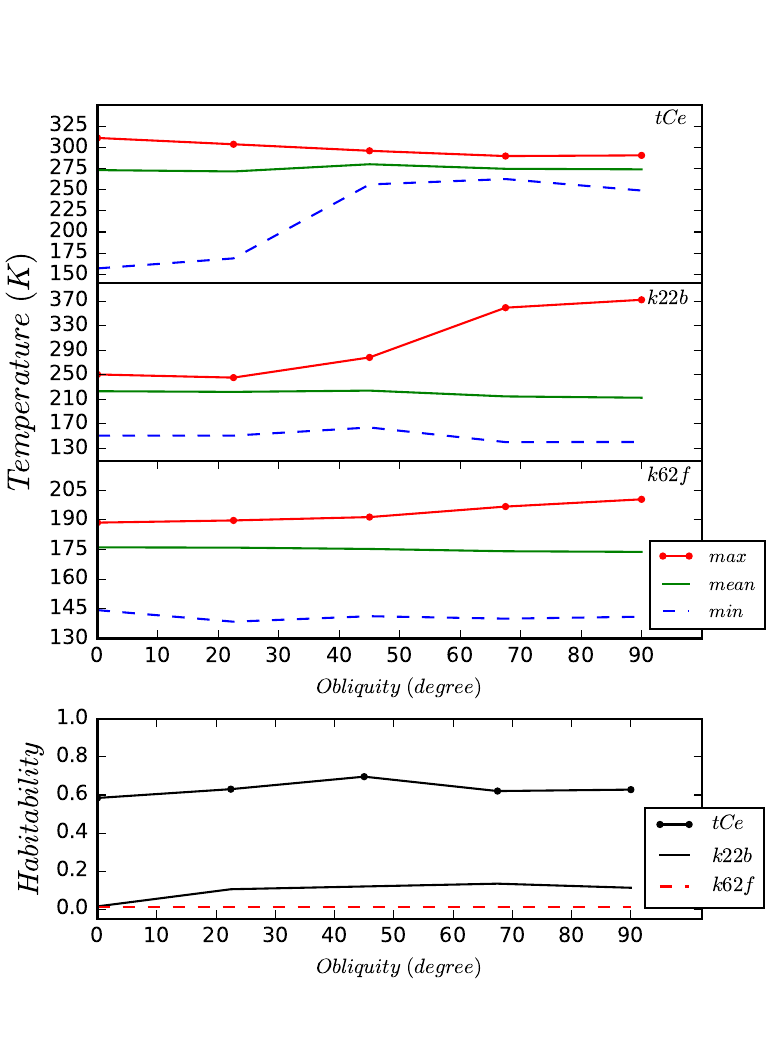}
        \caption{O.F. = 0.1, D.P. = 0.5 days, Pressure = $p_1$}
        \label{fig10-c}
    \end{subfigure}
 \hfill 
    \begin{subfigure}{0.49\textwidth}
        \includegraphics[width=\textwidth]{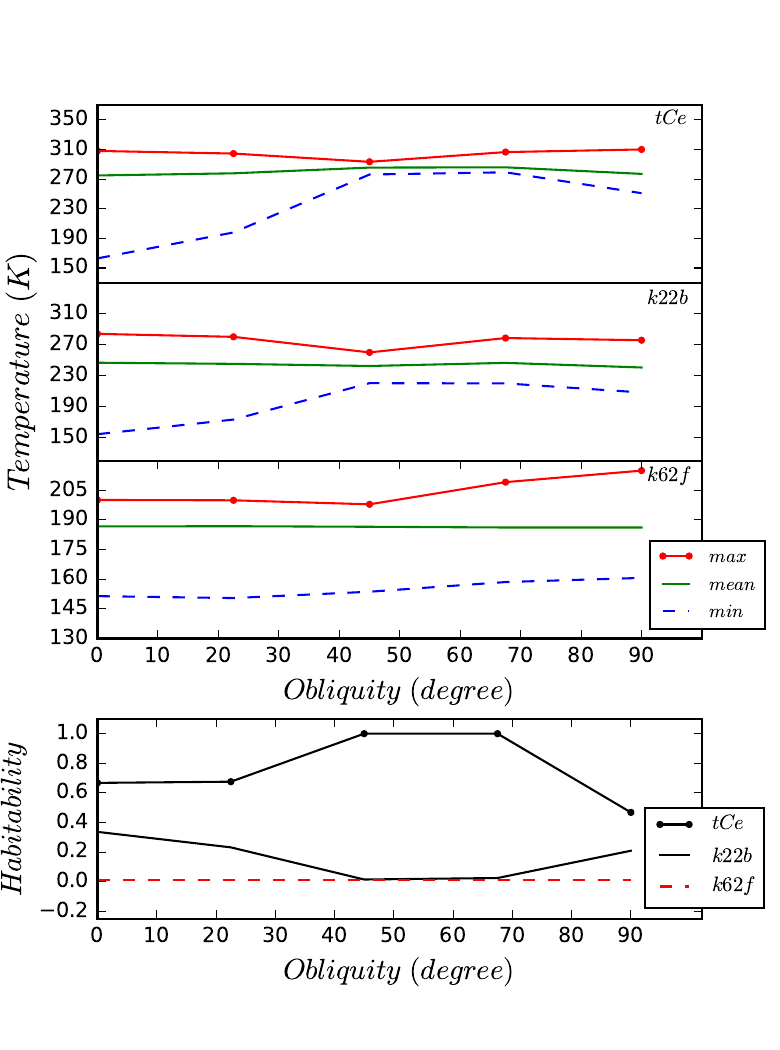}
        \caption{O.F. = 0.9, D.P. = 0.5 days, Pressure = $p_1$}
        \label{fig10-d}
    \end{subfigure}
    \caption{Spin obliquity varies; for planet $\tau$ Ceti e (tCe),  Kepler-22b (k22b) $p_1 =0.1\ atm, \text{\ }p_5 =1.3\ atm$; and for Kepler-62f (k62f) $p_1 =0.5\ atm, \text{\ }p_5 =4.5\ atm$ }
    \label{fig10}
\end{figure}

 When spin obliquity is zero, high latitudes receive a negligible amount of solar energy. Therefore, ice poles are created and spread into other latitudes until the received instellation compensates for the negative effect of albedo and OLR on the stored energy. By increasing the spin obliquity, the poles receive more instellation, which in turn decreases the area of ice caps. In the median obliquities around 45 degrees, since all latitudes receive a moderate amount of instellation, the temperature profile experiences less diversity. When this increment reaches more radical values such as 90 degrees, an ice belt is created on the equator since this is the region with the least instellation. However, the equator of a zero degress spin obliquity planet receives more yearly instellation in comparison to the poles of a 90 degrees spin obliquity planet. Therefore, the ice belt on the equator is narrower and warmer in general. If we compare the maximum, minimum, and mean temperatures of the cases in figure \ref{fig10}, the general trend of higher temperatures in 90 degrees spin obliquity compared with zero degrees obliquity is evident. Regarding the numerical difference between maximum and minimum temperatures for the planets at \ref{fig10-a}, this difference for $\tau$ Ceti e changes from 135 K at zero obliquity to 40 K at 45 degrees obliquity, and 82 K at 90 degrees obliquity. These values are 109, 56, and 128 Kelvin for Kepler-22b and 27, 13, and 22  Kelvin for Kepler-62f, respectively.

 
  The habitability fraction follows a more complicated process. When the planet has a relatively moderate or high temperature, adjusting the temperature profile to a more uniform one by increasing the obliquity causes the habitability to increase since there are more regions with a moderate temperature. Reaching more radical values of obliquity makes temperature profiles again more diverse with a higher density of frozen and boiling regions; therefore, habitability decreases. As an example, the habitability of $\tau$ Ceti e in the case of figure \ref{fig10-a} starts with a value of 0.52, reaches the value of 1.0 and then decreases again to the value of 0.77.
  
   When the planet has a relatively low temperature, a reverse process occurs. In low obliquities, parts of the planet which receive more instellation are habitable. When spin obliquity increases, it again makes the temperature profile more uniform. This uniformity results in lowering the temperature in above freezing regions and raising the temperature in frozen regions. However, as the instellation is not high enough the overall effect is to turn more regions into frozen rather than turning others to becoming more habitable. Therefore, habitability decreases. In high obliquities there is a similar situation to low obliquities; there are frozen parts with high instellation which begin to melt. These areas are responsible for increasing the habitability. The planet Kepler-22b in the case of figure \ref{fig10-d} is a good example, where its habitability starts from 0.34, decreases to zero, and then increases back to 0.22.
   
 There are some cases for which changing obliquity has no significant effect on habitability. These cases might be completely frozen in any spin condition or the dominant parameter is a parameter other than obliquity (e.g. pressure or diurnal period). Moreover, changing the status of some cases from transient to habitable or vice versa is considerable in figures \ref{fig5} and \ref{fig6}. By increasing obliquity, planet $\tau$ Ceti e experiences transformation from habitable to transient in low pressures and vice versa in high pressures. The planet Kepler-22b experiences transformation from habitable to transient in moderate pressures.
 

\nolinenumbers
\subsection{Changing the Diurnal Period}

  As mentioned earlier, due to the limitation of the model, the diurnal period should be small enough with respect to the orbital period. According to equation \ref{r7}, increasing the rotational velocity results in decreasing the diffusion coefficient. This means that heat cannot transfer effectively between different latitudes. This is a result of the combination of the facts that as rotational rate increases, the decreases in eddy length scale leads to less eddy transport poleward. On the contrary, as the rotation slows down, large planetary scale Hadley cells increase the heat transport by the mean meridional circulation (Kaspi \& Showman, 2015). In fact, we expect that the main impact of increasing the diurnal period should be in changing the latitudinal temperature profile where the difference between the maximum and minimum temperatures decreases. However, the effect of changing this parameter on mean temperature is negligible.
   
   Since increasing the diurnal period converges the maximum and minimum temperatures, if there is an effect on habitability, that effect would be increasing the habitability as a result of making hotter regions colder and colder regions warmer, figure \ref{zaki}. For instance, in figure \ref{fig11-b},  the difference between maximum and minimum temperatures of the planet $\tau$ Ceti e starts with a value of 134 K and finally reaches 31 Kelvin. The two other planets follow nearly the same process. The habitability for $\tau$ Ceti e increases from 0.53 to 1.0.


\nolinenumbers
\subsection{Statistical interface to the unknown parameters}
\label{sec5}

Up until now, we investigated the signature of the mentioned unknown parameters on the habitability of an exoplanet. As we see, they have a variety of effects on the temperature and habitability and should be taken into account carefully. The problem with these parameters is that we just know or even suspect a possible range for them but not their exact values. Therefore, to quantify their effect we utilized a statistical approach.

   With the values for the parameters introduced in section \ref{lebm}, the simulation's phase space contains 625 points. To have a better perception of how the temperature and habitability of a planet vary when a selected parameter or a combination of parameters changes, we encapsulated the result of all of the diagrams in one figure called ``Map of parameters''. There are three of these figures (figures \ref{fig5} to \ref{fig7}), one for each of the three planets. In these figures, by moving to the right, the pressure increases in each little square and the obliquity increases by jumping between the large squares. Similarly, by moving to the top, the diurnal period increases in each little square and the ocean fraction increases by jumping between the large squares. The habitability status and mean temperature of each case could be determined using the guidance next to each figure.

   From these maps a planet could have different states based on the value of its four unknown parameters. Therefore, setting these values to those of the Earth does not give us an accurate understanding of the habitability of an exoplanet. For instance, figure  \ref{fig6} shows that 333 out of 625 cases (slightly more than 50 percent) have a habitability more than 0.4. However, it is still an inaccurate inference to say that the planet is likely to be habitable rather than unhabitable.
   
To assess this issue more accurately, we need to meticulously determine our phase space. For the spin obliquity, the possible range is covered in the simulation. Ocean fraction takes the values from 0 to 1 with the exception of 0.0 since the planet with no water is unhabitable in our assumption of habitability. The pressure could not be 0.0 atm which indicates no atmosphere on the planet and should be excluded from the possible range. The upper limit of 1,100 atm for the pressure is a good choice based on the endurance of extremophiles in an extreme environment (Stan-Lotter, 2007). To make a statistical interpretation and comparison, it is essential to specify a single phase space for all three planets. In this regard, the phase space on the pressure coordinate will be confined to the interval (0.0, 1]. Choosing this interval helps us avoid boundaries problem during the extrapolations. Moreover, it is a representation of different atmospheric pressures on the Earth. The value of diurnal period depends on the formation history of the planet and has no apparent bounds. However, using the statistical approach based on a simulation of planet formation with different initial planetary system parameters it is possible to determine the distribution of the diurnal period (Miguel \& Brunini, 2010). According to those results, the most probable values for diurnal period are larger than 10 hours but the number of planets with diurnal periods between 0.1 and 10 hours is also considerable. Values less than 0.1 hours are really rare, because at those periods the spin angular velocities exceed the critical  instability angular velocity value. Therefore, we set the lower limit of 0.1 hours for the diurnal period. The upper limit could not be bounded. Using the probability distribution for a primordial diurnal period introduced in the aforementioned study we can discuss the habitability probability with the help of conditional probability by computing probability of habitability given that the probability of diurnal period lies in a specific range; this is the only way to reconcile the probability and LEBM inherent limitations on diurnal period.
   
   Considering the above discussion, we set the range (0.0 - 1.0] atm for pressure, (0.0 - 1.0] for ocean fraction, [0.0 - 90.0] degrees for spin obliquity, and (0.0042 - 1.5] days for diurnal period. Using Miguel \& Brunini (2010) the probability of diurnal period being within this mentioned range can be calculated and is equal to 0.1226.
   
   Since the parameters are continuous, to calculate the probabilities we need the area of the region which gives us the desired outcome. Running the simulation for each point in phase space is not possible. To overcome the situation we used the Monte Carlo method. We generated many random coordinates in the domain of possible values. Then, we used a weighted average according to the simulated point to estimate the habitable fraction and habitability status.
   
   To estimate those values in the 4D space of the parameters by using a weighted average, we first determine the nearest neighbor simulated coordinate and the 4D cube in which the random value is located. Then, we calculated the average according to this equation:
   
\begin{equation}
\text{avg}(v) = \frac{\sum\limits_{xyzw}v_{xyzw} e^{-a d_{xyzw}}}{\sum\limits_{xyzw}e^{-a d_{xyzw}}}
\label{ave}
\end{equation}

where $v_{xyzw}$ is the simulated value for each corner of the 4D cube and $d_{xyzw}$ is the normalized-to-step-size Euclidean distance from each nearest neighbors:
\begin{equation}
\begin{split}
d_{xyzw} =& \Big(\big(\frac{P-P_{min}}{P_{max}-P_{min}}\big)^2+\big(\frac{O.F.-O.F._{min}}{O.F._{max}-O.F._{min}}\big)^2+
 \big(\frac{O.b.-O.b._{min}}{O.b._{max}-O.b._{min}}\big)^2 + \big(\frac{T-T_{min}}{T_{max}-T_{min}}\big)^2 \Big)^{\frac{1}{2}}
\end{split}
\end{equation}
The coefficient $a$ is set to be the Napierian logarithm of $2.220446\times 10^{-16}$ (the smallest positive number in floating-point representation with double precision). With this value for a distance equal to the edge of the cube, the exponential value becomes technically zero in the calculation; if the random value is located on the vertices, the average yields the simulated value.

   Habitability fraction is a continuous variable which we can estimate using equation \ref{ave}. Habitability status, on the other hand, is a categorical variable; to average it, we assign +1, 0, and -1 respectively for habitable, transient and unhabitable status. The averaged value will be rounded to the nearest category number.

  There are 30000 generated random points which is high enough such that repeating the Monte Carlo process changes the value of outcomes by less than 0.1\%.  The results for our three surveyed planets are summarized in table \ref{probability_table}. To calculate the probability of habitability status (habitable - transient - unhabitable) we need to multiply the habitability probability given that the diurnal period lies in the range of (0.0042 - 1.5] by the probability of the diurnal period being in the same range.

\begin{table}
	\begin{center}
		\caption{Conditional probability integrated over all planetary parameters studied using the Monte Carlo approach}
		\begin{tabular}{|c | c | c | c |}
			\hline
			value & $\tau$ Ceti e & Kepler-22b & Kepler-62f  \\ \hline
			Habitable fraction  & 0.836 & 0.408 & 0.0  \\ 
			Transient fraction  & 0.121 & 0.419 & 0.0  \\ 
			Unhabitable fraction & 0.043 & 0.173 & 1.0 \\
			\hline
		\end{tabular}
		\label{probability_table}
	\end{center}
\end{table}



\nolinenumbers
\section{Conclusion}
\label{conclusion}

\begin{figure}[H]
    \centering
    \begin{subfigure}{0.49\textwidth}
        \includegraphics[width=\textwidth]{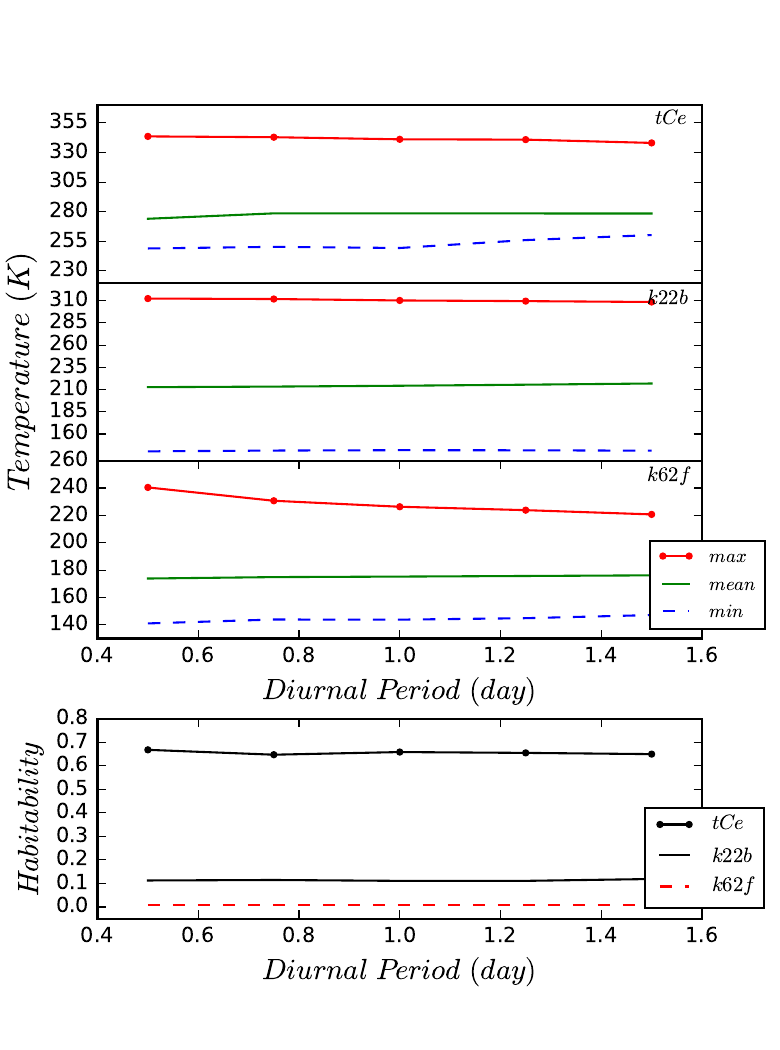}
        \caption{O.F. = 0.1, S.O. = 90.0 degrees, Pressure = $p_1$}
        \label{fig11-a}
    \end{subfigure}
 \hfill 
    \begin{subfigure}{0.49\textwidth}
        \includegraphics[width=\textwidth]{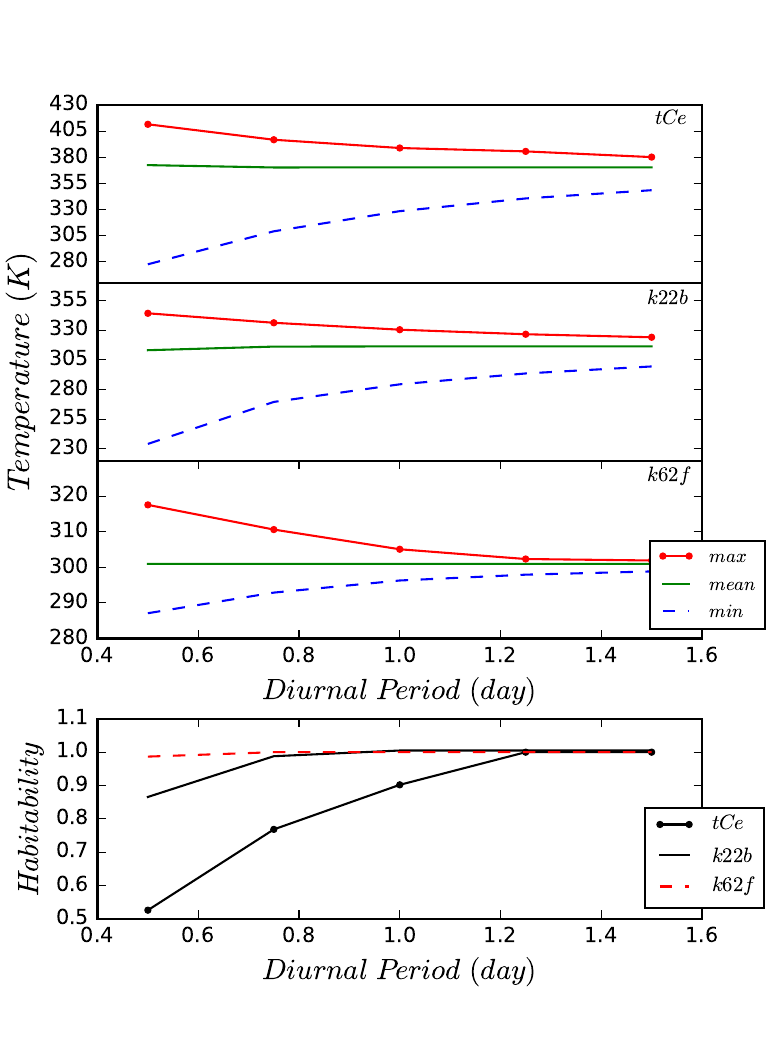}
        \caption{O.F. = 0.1,  S.O. = 0.0 degrees, Pressure = $p_5$}
        \label{fig11-b}
    \end{subfigure}
    \newline
 \hfill 
    \begin{subfigure}{0.49\textwidth}
        \includegraphics[width=\textwidth]{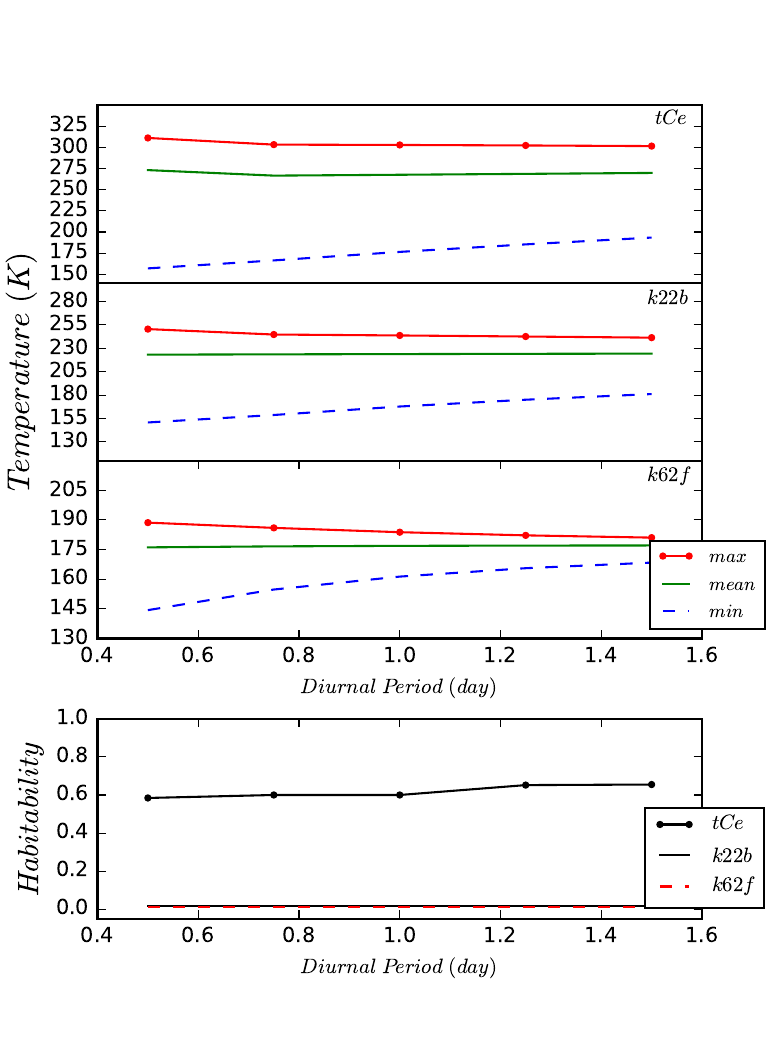}
        \caption{O.F. = 0.1, S.O. = 0.0 degrees, Pressure = $p_1$}
        \label{fig11-c}
    \end{subfigure}
 \hfill 
    \begin{subfigure}{0.49\textwidth}
        \includegraphics[width=\textwidth]{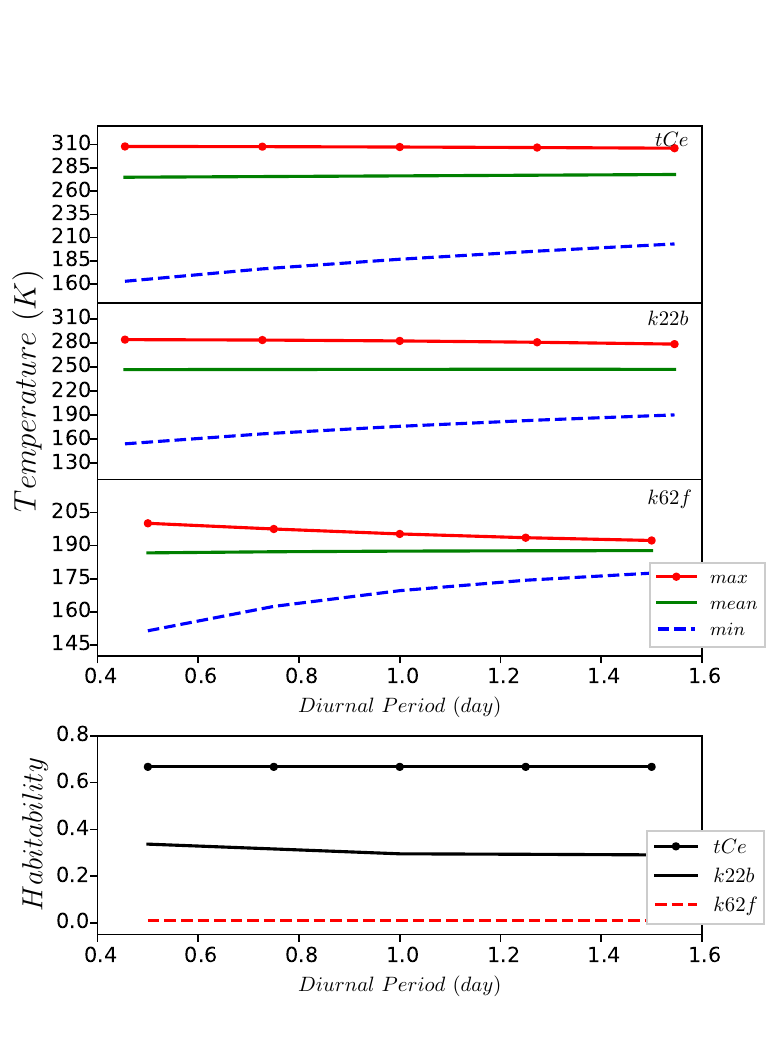}
        \caption{O.F. = 0.9, S.O. = 0.0 degrees, Pressure = $p_1$}
        \label{fig11-d}
    \end{subfigure}
    \caption{Diurnal period varies; for planet $\tau$ Ceti e (tCe), Kepler-22b (k22b) $p_1 =0.1\ atm, \text{\ }p_5 =1.3\ atm$; and for Kepler-62f (k62f) $p_1 =0.5\ atm, \text{\ }p_5 =4.5\ atm$ }
    \label{fig11}
\end{figure}

We applied a latitudinal energy balance model to study the effect of unknown parameters of an exoplanet which cannot be obtained by observation. These parameters include surface pressure (P), ocean fraction (O.F.), spin obliquity (O.b.), and diurnal period (T). Starting with the diffusion equation \ref{r1}, we tested its reliability by applying it to the Earth data (See Appendix \ref{sectionthree}). Comparing the result of our simulation with the data of NCEP/NCAR (National Centers for Environmental Prediction/ National Center for Atmospheric Research) (North \& Coakley, 1979) global temperature data shows that the model, despite its simplicity, is precise enough to make it suitable for our survey. The main improvement of our simulation compared with Forgan (2013) and Spiegel (2009) is defining some parameters based on Vladilo et. al. (2013). When the albedo is defined by relating it only to the temperature, a global freezing problem occurs which is dependent on the starting point of the planet in its orbit. For instance, by increasing the semi-major axis of the Earth to 1.025 AU, based on the defined starting point, the whole surface of the Earth could turn into ice. This problem is solved when equation \ref{r8}, which uses the albedo of land, ocean, ice, and cloud portions on each of those surfaces, is applied (see Appendix \ref{AppNewAlbido}). To study the impact of each parameter on the temperature and habitability, three maps of parameters for each exoplanet consisting of 625 different sets of parameters were formed. The effect of a parameter is shown in four different sets of parameters extracted from the maps in which one parameter is free, changing from a minimum to a maximum value. These effects are shown in figures \ref{fig8} to  \ref{fig11} .

   The pressure has a strong, nearly linear, impact on temperature. It is the dominant parameter in the examined range of our simulations. When it is increased, the temperature of the planet is also increased. Habitability, on the other hand, could undergo different scenarios. Depending on the mean temperature of the planet at the starting point, it could increase when more frozen regions start to melt or decrease when more regions go beyond the boiling point of water.

  Since the ocean has a much higher thermal inertia, the effect of increasing its ratio is to increase the total heat capacity of the surface. In a short period, it prevents abrupt changes in the temperature; however, over the long term, this feature does not play a role as the planet reaches a stable condition. The other impact of increasing the ocean fraction is reducing the difference between the maximum and minimum temperatures. These effects make the planet more habitable by increasing the habitability fraction or changing the status of the planet from transient to habitable.
  
   Spin obliquity is changed from 0 to 90 degrees. In low obliquities, ice caps are created and therefore the difference between the maximum and minimum temperatures is high. As the obliquity increases, ice caps start to melt and these extreme temperatures approach each other. In very high obliquities, where equators receive the least instellation, ice belts are formed on the equator and thus the difference between the maximum and minimum temperatures starts to increase.
   
  The main impact of the increasing diurnal period is a better heat transfer between latitudes which in turn decreases the difference between the maximum and minimum temperatures of the planet. The value of the diurnal period does not have a noticeable effect on the mean temperature, and so the habitability does not change by varying this parameter.
   
   Since the LEBM does not depend on the mass and radius of a planet, it is expected that nearly the same results for other exoplanets located in the inner, middle, and outer regions of the HZ would be obtained. However, it should be mentioned that the radius and gravity of a planet could have a significant effect on its climate for an even broader range. These effects are shown by Kopparapu et al. (2014), Kaspi \& Showman (2015), and Yang et al. (2019).

\begin{figure}[H]
	\begin{center}
		\includegraphics[scale=1.8]{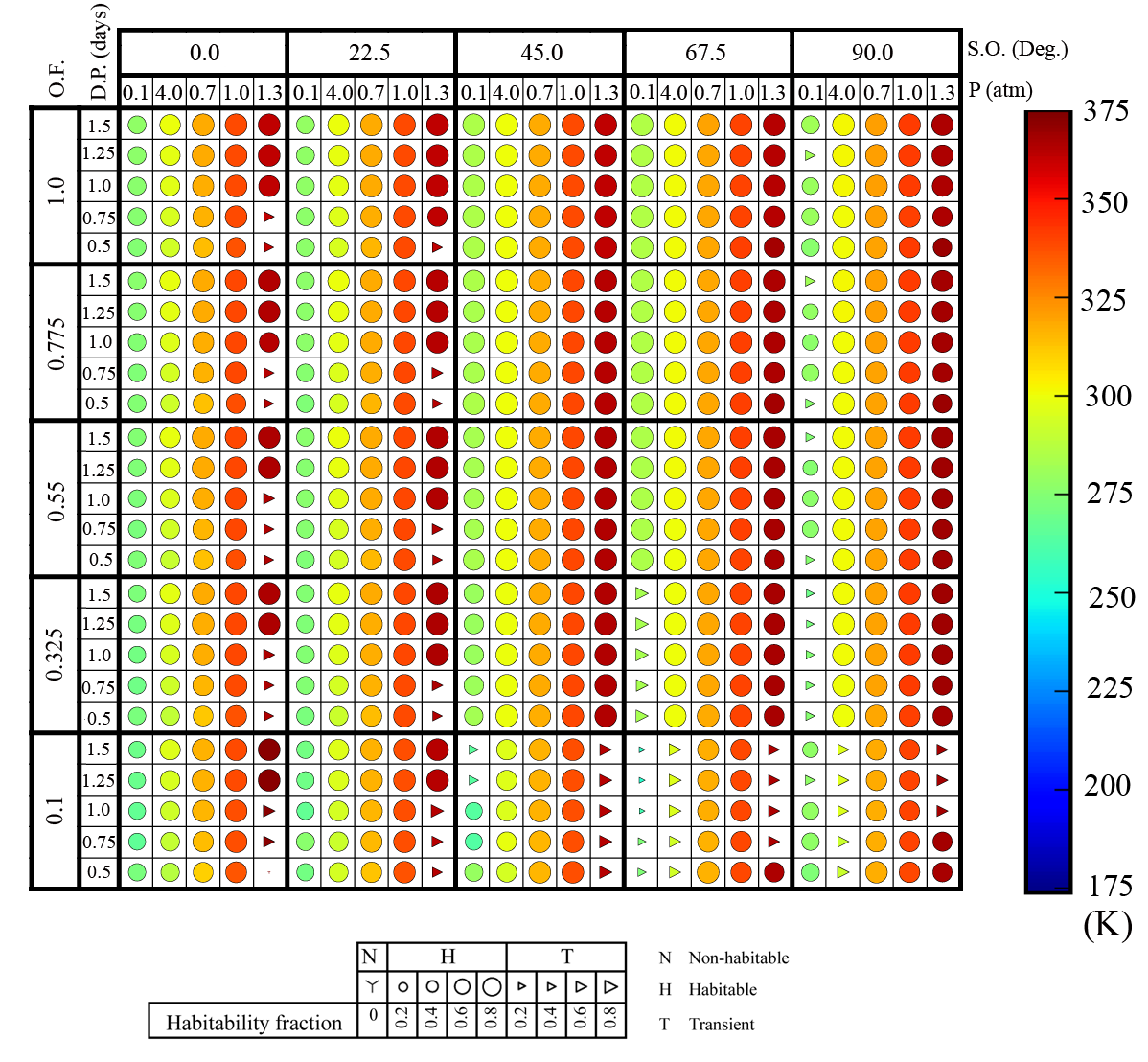}
	\end{center}
	
	\caption{$\tau$ Ceti e parameters map - All parameters are in an increasing order from left to right and bottom to up. Pressures are (0.1, 0.4, 0.7, 1, 1.3) atm, Diurnal Periods are (0.5, 0.75, 1.0, 1.25, 1.5) days, Spin Obliquities are (0.0, 22.5, 45.0, 67.5, 90.0) degrees, and Ocean Fractions are (0.1, 0.325, 0.55, 0.775, 1.0). They are displayed in the map, too.}
	\label{fig5}
\end{figure}

\begin{figure}
	\begin{center}
		\includegraphics[scale=1.8]{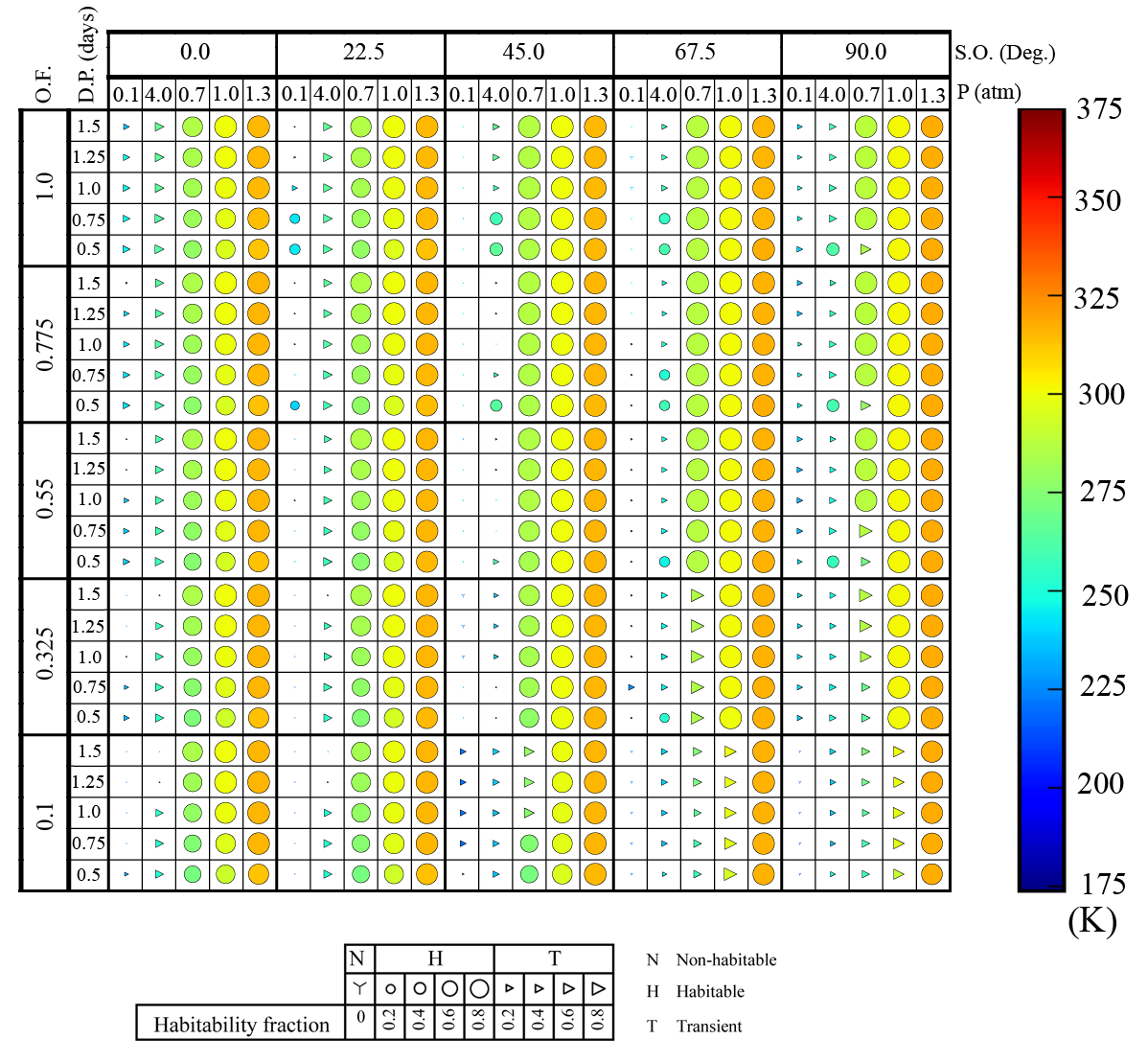}
	\end{center}
	
	\caption{Kepler-22b parameters map - All parameters are in an increasing order from left to right and bottom to up. Pressures are (0.1, 0.4, 0.7, 1, 1.3) atm, Diurnal Periods are (0.5, 0.75, 1.0, 1.25, 1.5) days, Spin Obliquities are (0.0, 22.5, 45.0, 67.5, 90.0) degrees, and Ocean Fractions are (0.1, 0.325, 0.55, 0.775, 1.0). They are displayed in the map, too.}
	\label{fig6}
\end{figure}

\begin{figure}
	\begin{center}
		\includegraphics[scale=1.8]{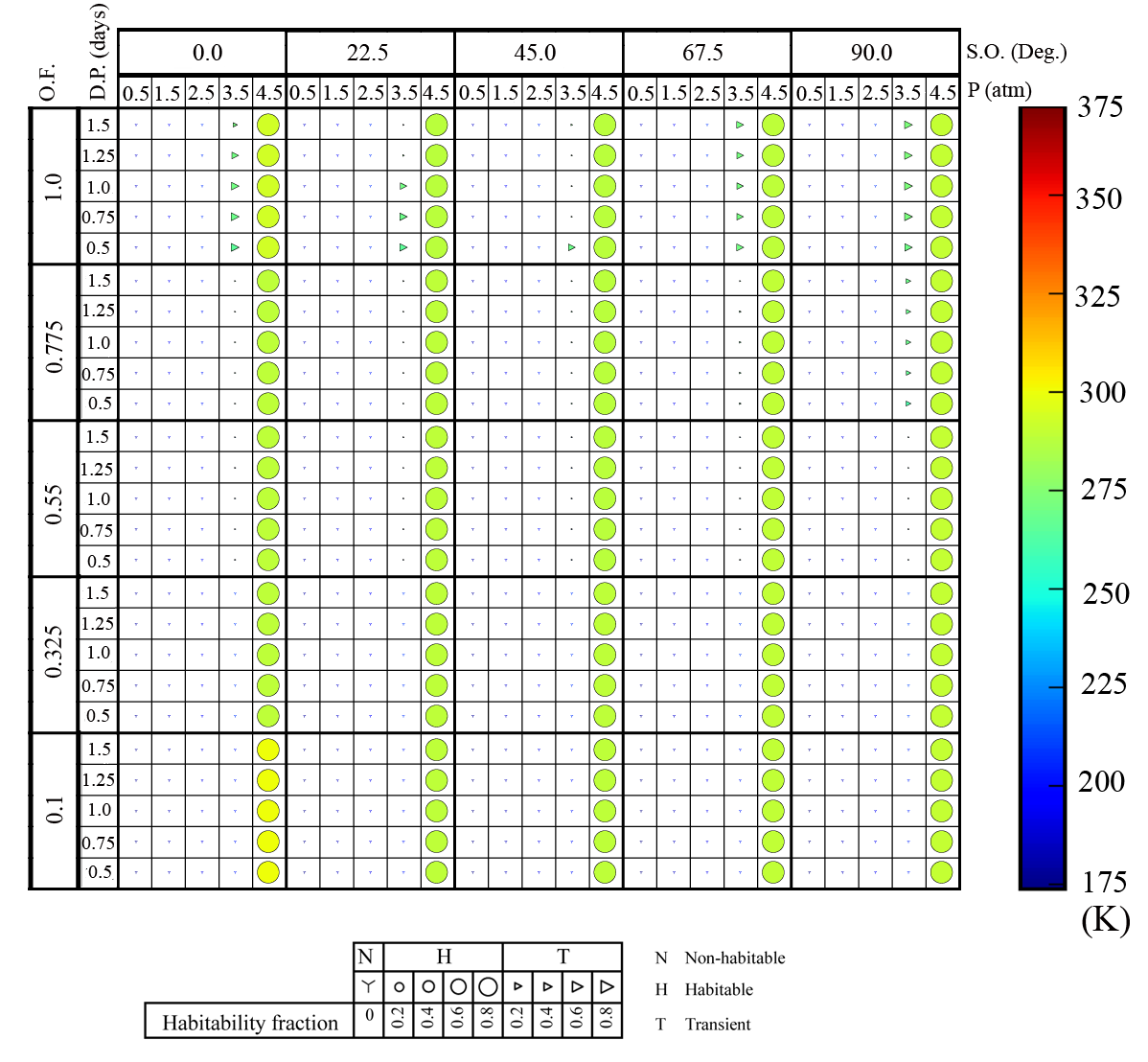}
	\end{center}
	
	\caption{Kepler-62f parameters map - 
		 Kepler-62f parameters map - All parameters are in an increasing order from left to right and bottom to up. Pressures are (0.1, 0.4, 0.7, 1, 1.3) atm, Diurnal Periods are (0.5, 0.75, 1.0, 1.25, 1.5) days, Spin Obliquities are (0.0, 22.5, 45.0, 67.5, 90.0) degrees, and Ocean Fractions are (0.1, 0.325, 0.55, 0.775, 1.0). They are displayed in the map, too.}
	\label{fig7}
\end{figure}

\begin{center}
\begin{figure}[H]
	\centering
	\includegraphics[width=0.7\textwidth]{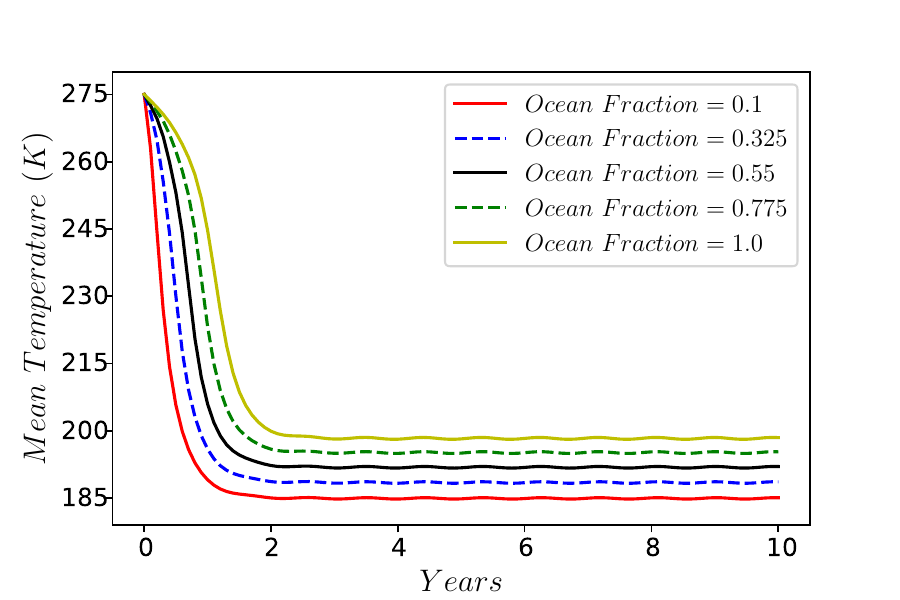}
	\caption{Simulation for a Sun-Earth system with different ocean fraction which compares the required time to reach a semi-stable condition}
	\label{last}
\end{figure}
\end{center}

\begin{center}
\begin{figure}[H]
	\centering
	\includegraphics[width=0.7\textwidth]{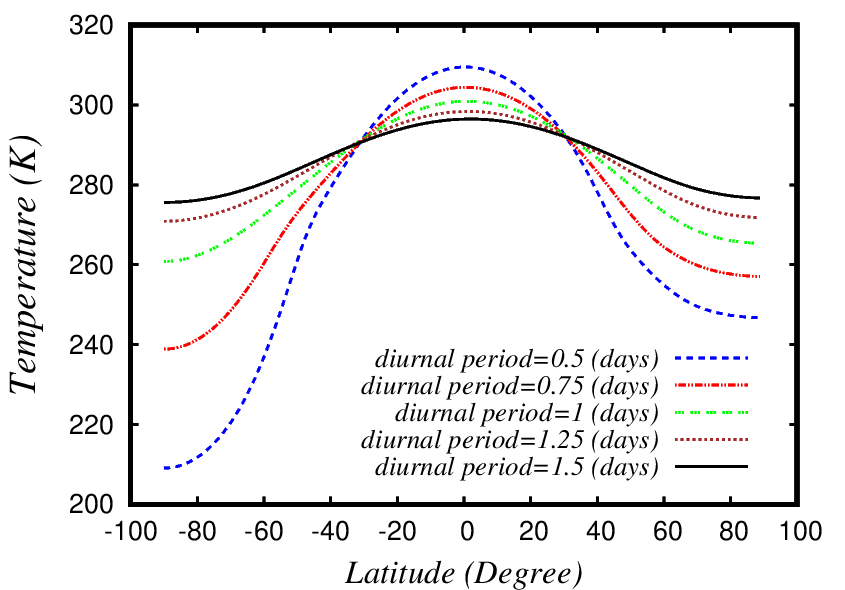}
	\caption{Simulation for a Sun-Earth system with different diurnal period of the Earth shows that increasing this parameter results in a more homogeneous temperature distribution}
	\label{zaki}
\end{figure}
\end{center}

   A better understanding of the changing of unknown parameters simultaneously could be achieved by probing the ``Map of Parameters'' for each planet in figures  \ref{fig5} to \ref{fig7}. To make any statistical inference from these maps, we need to have the area of the phase space corresponding to our desired outcome. Utilizing a Monte Carlo process, we created a large number of points in the phase space and estimated the value of each point with a weighted averaging method. This approach is inevitable since the random value is continuous and it is not possible to run a simulation for each point. 
   
   As mentioned in section \ref{discussion}, there is no upper bound for the diurnal period. Thus, we utilized conditional probability using Miguel \& Brunini (2010). We set the probability of the diurnal period to be in the range of (0.0042 - 1.5] days, and calculated the conditional probability of habitability of the planet, having the ranges of (0.0 - 1.0] atm for pressure, (0.0 - 1.0] for ocean fraction, [0.0 - 90.0] degrees for spin obliquity, given that the diurnal period is in the aforementioned range. To calculate the probability of habitability status we needed to multiply the conditional probability with the probability of diurnal period being in that specific range.

For each planet, we simulated 30000 points of phase space for the diurnal period to be in the range of (0.0042 - 1.5] days, having the ranges of (0.0 - 1.0] atm for pressure, (0.0 - 1.0] for ocean fraction, and [0.0 - 90.0] degrees for spin obliquity. Then, we computed the conditional probability habitability status given that the diurnal period is the in range of (0.0042 - 1.5] days. As it is presented in table II, $\tau$ Ceti e is mostly habitable, the probability of Kepler-22b to be habitable and transient is almost equal, and Kepler-62 is unhabitable in those aforementioned ranges.

\nolinenumbers
\section{Acknowledgement}

The project is supported by the Eclipse Department of IOTA/Middle East. We like to express our sincere thanks to Paul D. Maley for his cooperation in editing the text. Also, we thank the referee for his/her insightful comments and remarks to improve the manuscript.
\nolinenumbers
\section{references}
\begin{enumerate}

\item Akeson, R.L., Chen, X., Ciardi, D., Crane, M., Good, J., Harbut, M., Jackson, E., Kane, S.R., Laity, A.C., Leifer, S. and Lynn, M., 2013. The NASA exoplanet archive: data and tools for exoplanet research. \textit{Publications of the Astronomical Society of the Pacific}, 125(930), p.989.
\item Borucki, W.J., Agol, E., Fressin, F., Kaltenegger, L., Rowe, J., Isaacson, H., Fischer, D., Batalha, N., Lissauer, J.J., Marcy, G.W. and Fabrycky, D., 2013. Kepler-62: a five-planet system with planets of 1.4 and 1.6 Earth radii in the habitable zone.  \textit{Science}, p.1234702.
\item Borucki, W. J., Koch, D. G., Basri, G., Batalha, N., Brown, T. M., Bryson, S. T., \& Dunham, E. W. (2011). Characteristics of planetary candidates observed by Kepler. II. Analysis of the first four months of data.  \textit{The Astrophysical Journal}, 736(1), 19. [KOI-87.01]
\item Borucki, W.J., Koch, D.G., Batalha, N., Bryson, S.T., Rowe, J., Fressin, F., Torres, G. Caldwell, D.A., Christensen-Dalsgaard, J., Cochran, W.D. and DeVore, E., 2012. Kepler-22b: A 2.4 Earthradius planet in the habitable zone of a Sun-like star.  \textit{The Astrophysical Journal}, 745(2), p.120.
\item Brown, T.M., Latham, D.W., Everett, M.E. and Esquerdo, G.A., 2011. Kepler input catalog: photometric calibration and stellar classifcation.  \textit{The Astronomical Journal}, 142(4), p.112.
\item Buchhave, L.A., Bizzarro, M., Latham, D.W., Sasselov, D., Cochran, W.D., Endl, M., Isaacson, H., Juncher, D. and Marcy, G.W., 2014. Three regimes of extrasolar planet radius inferred from host star metallicities.  \textit{Nature}, 509(7502), p.593.
\item Budyko, M.I., 1969. The effect of solar radiation variations on the climate of the earth.  \textit{tellus}, 21(5), pp.611-619.
\item Cess, R.D., 1976. Climate change: An appraisal of atmospheric feedback mechanisms employing zonal climatology. \textit{Journal of the Atmospheric Sciences}, 33(10), pp.1831-1843.
\item Charnay, B., Forget, F., Wordsworth, R., Leconte, J., Millour, E., Codron, F. and Spiga, A., 2013. Exploring the faint young Sun problem and the possible climates of the Archean Earth with a 3‐D GCM. \textit{Journal of Geophysical Research: Atmospheres}, 118(18), pp.10-414.
\item Chemke, R., Kaspi, Y., Halev, I. 2016. The thermodynamic effect of atmospheric mass on early Earth’s temperature, \textit{Geophys. Res. Lett.}, 43, pp11,414–11,422.
\item Colose, C.M., Del Genio, A.D. and Way, M.J., 2019. Enhanced habitability on high obliquity bodies near the outer edge of the habitable zone of Sun-like stars. \textit{The Astrophysical Journal}, 884(2), p.138.
\item Feng, F., Tuomi, M., Jones, H.R., Barnes, J., Anglada-Escude, G., Vogt, S.S. and Butler, R.P., 2017. Color difference makes a difference: Four planet candidates around $\tau$ Ceti.  \textit{The Astronomical Journal}, 154(4), p.135.
\item Fischer, D.A., Howard, A.W., Laughlin, G.P., Macintosh, B., Mahadevan, S., Sahlmann, J. and Yee, J.C., 2015. Exoplanet detection techniques.  \textit{arXiv preprint arXiv:1505.06869}.
\item Forgan, D., 2013. Assessing circumbinary habitable zones using latitudinal energy balance modelling.  \textit{Monthly Notices of the Royal Astronomical Society}, 437(2), pp.1352-1361.
\item Franck, S., Von Bloh, W., Bounama, C., Steffen, M., Schönberner, D. and Schellnhuber, H.J., 2000. Determination of habitable zones in extrasolar planetary systems: Where are Gaia’s sisters?.  \textit{Journal of Geophysical Research: Planets}, 105(E1), pp.1651-1658.
\item Gaidos, E. and Williams, D.M., 2004. Seasonality on terrestrial extrasolar planets: inferring obliquity and surface conditions from infrared light curves.  \textit{New Astronomy}, 10(1), pp.67-77.
\item Goldblatt, C., Claire, M.W., Lenton, T.M., Matthews, A.J., Watson, A.J. and Zahnle, K.J., 2009. Nitrogen-enhanced greenhouse warming on early Earth. \textit{Nature Geoscience}, 2(12), pp.891-896.
\item Heng, K., \& Showman, A. P. (2015). Atmospheric dynamics of hot exoplanets.  \textit{Annual Review of Earth and Planetary Sciences}, 43, 509-540. 
\item Keles, E., Grenfell, J.L., Godolt, M., Stracke, B. and Rauer, H., 2018. The Effect of Varying Atmospheric Pressure upon Habitability and Biosignatures of Earth-like Planets. \textit{Astrobiology}, 18(2), pp.116-132.
\item Kane, S.R., Hill, M.L., Kasting, J.F., Kopparapu, R.K., Quintana, E.V., Barclay, T., Batalha, N.M., Borucki, W.J., Ciardi, D.R., Haghighipour, N. and Hinkel, N.R., 2016. A catalog of Kepler habitable zone exoplanet candidates.  \textit{The Astrophysical Journal}, 830(1), p.1.
\item Kaspi, Y. and Showman, A.P., 2015. Atmospheric dynamics of terrestrial exoplanets over a wide range of orbital and atmospheric parameters.  \textit{The Astrophysical Journal}, 804(1), p.60.
\item Kasting, J.F., Whitmire, D.P. and Reynolds, R.T., 1993. Habitable zones around main sequence stars.  \textit{Icarus}, 101(1), pp.108-128.
\item Kopparapu, R.K., 2013. A revised estimate of the occurrence rate of terrestrial planets in the habitable zones around Kepler M-dwarfs.  \textit{The Astrophysical Journal Letters}, 767(1), p.L8.
\item Kopparapu, R.K., Ramirez, R., Kasting, J.F., Eymet, V., Robinson, T.D., Mahadevan, S., Terrien, R.C., Domagal-Goldman, S., Meadows, V. and Deshpande, R., 2013. Habitable zones around main-sequence stars: new estimates. \textit{The Astrophysical Journal}, 765(2), p.131.
\item Kopparapu, R.K., Ramirez, R.M., SchottelKotte, J., Kasting, J.F., Domagal-Goldman, S. and Eymet, V., 2014. Habitable zones around main-sequence stars: dependence on planetary mass. \textit{The Astrophysical Journal Letters}, 787(2), p.L29.
\item kumar Kopparapu, R., Wolf, E.T., Arney, G., Batalha, N.E., Haqq-Misra, J., Grimm, S.L. and Heng, K., 2017. Habitable moist atmospheres on terrestrial planets near the inner edge of the habitable zone around M dwarfs. \textit{The Astrophysical Journal}, 845(1), p.5.
\item kumar Kopparapu, R., Wolf, E.T., Haqq-Misra, J., Yang, J., Kasting, J.F., Meadows, V., Terrien, R. and Mahadevan, S., 2016. The inner edge of the habitable zone for synchronously rotating planets around low-mass stars using general circulation models. \textit{The Astrophysical Journal}, 819(1), p.84.
\item Mayor, M. and Queloz, D., 1995. A Jupiter-mass companion to a solar-type star.  \textit{Nature}, 378(6555), p.355.
\item Miguel, Y. and Brunini, A., 2010. Planet formation: statistics of spin rates and obliquities of extrasolar planets.  \textit{Monthly Notices of the Royal Astronomical Society}, 406(3), pp.1935-1943.
\item North, G.R. and Coakley, J.A., 1979. a Seasonal Climate Model for Earth. \textit{In Evolution of Planetary Atmospheres and Climatology of the Earth} (p. 249).
\item Pierrehumbert, R.T., 2010. \textit{Principles of planetary climate}. Cambridge University Press.
\item Rushby, A.J., Shields, A.L. and Joshi, M., 2019. The Effect of Land Fraction and Host Star Spectral Energy Distribution on the Planetary Albedo of Terrestrial Worlds. \textit{The Astrophysical Journal}, 887(1), p.29.
\item Salby, M.L., 2012. \textit{Physics of the Atmosphere and Climate}. Cambridge University Press.
\item Sellers, W. D. (1969). A global climatic model based on the energy balance of the earth atmosphere system. \textit{Journal of Applied Meteorology}, 8(3), 392-400.
\item Sellers, W.D., 2011. A global climatic model based on the energy balance of the earth atmosphere system. \textit{The Warming Papers}, 125.
\item Spiegel, D.S., Menou, K. and Scharf, C.A., 2008. Habitable climates. \textit{The Astrophysical Journal}, 681(2), p.1609.
\item Spiegel, D.S., Menou, K. and Scharf, C.A., 2009. Habitable climates: the influence of
obliquity. \textit{The Astrophysical Journal}, 691(1), p.596.
\item Teixeira, T.C., Kjeldsen, H., Bedding, T.R., Bouchy, F., Christensen-Dalsgaard, J., Cunha, M.S., Dall, T., Frandsen, S., Karoff, C., Monteiro, M.J.P.F.G. and Pijpers, F.P., 2009. Solar-like oscillations in the G8 V star $\tau$ Ceti. \textit{Astronomy \& Astrophysics}, 494(1), pp.237-242.
\item Torres, G., Kipping, D.M., Fressin, F., Caldwell, D.A., Twicken, J.D., Ballard, S. Batalha, N.M., Bryson, S.T., Ciardi, D.R., Henze, C.E. and Howell, S.B., 2015. Validation of 12 small Kepler transiting planets in the habitable zone. \textit{The Astrophysical Journal}, 800(2), p.99.
\item Vladilo, G., Murante, G., Silva, L., Provenzale, A., Ferri, G. and Ragazzini, G., 2013. The habitable zone of earth-like planets with different levels of atmospheric pressure.\textit{ The Astrophysical Journal}, 767(1), p.65.
\item Williams, D.M. and Kasting, J.F., 1997. Habitable planets with high obliquities. \textit{Icarus}, 129(1), pp.254-267.
\item Wolf, E. T., Toon O. B..  2014, Delayed onset of runaway and moist greenhouse climates for Earth, \textit{Geophys. Res. Lett.}, 41, pp167–172.
\item Yang, J., Boué, G., Fabrycky, D.C. and Abbot, D.S., 2014. Strong dependence of the inner edge of the habitable zone on planetary rotation rate. \textit{The Astrophysical journal letters}, 787(1), p.L2.
\item Yang, J., Cowan, N.B. and Abbot, D.S., 2013. Stabilizing cloud feedback dramatically expands the habitable zone of tidally locked planets. \textit{The Astrophysical Journal Letters}, 771(2), p.L45.
\item Zsom, A., Seager, S., De Wit, J. and Stamenković, V., 2013. Toward the minimum inner edge distance of the habitable zone. \textit{The Astrophysical Journal}, 778(2), p.109.
\end{enumerate}

\newpage
    
\newpage

\vspace{-0.9cm} 

\vspace{-0.9cm}

\newpage
\vspace{-0.9cm}

\vspace{-0.9cm}


The effect of changing the simulation's starting point and the planet-star distance in an Earth-Sun like system
 \appendix
\section{The effect of changing the simulation's starting point and the planet-star distance in an Earth-Sun like system} \label{AppNewAlbido}
\begin{figure}[]
    \centering
        \begin{subfigure}{0.49\textwidth}
        \includegraphics[width=\textwidth]{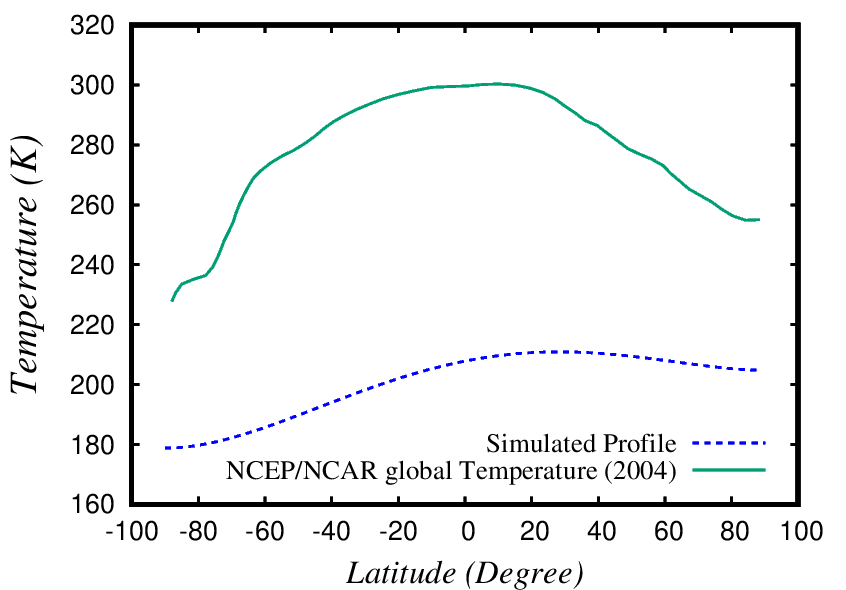}
    \end{subfigure} 
          \begin{subfigure}{0.49\textwidth}
        \includegraphics[width=\textwidth]{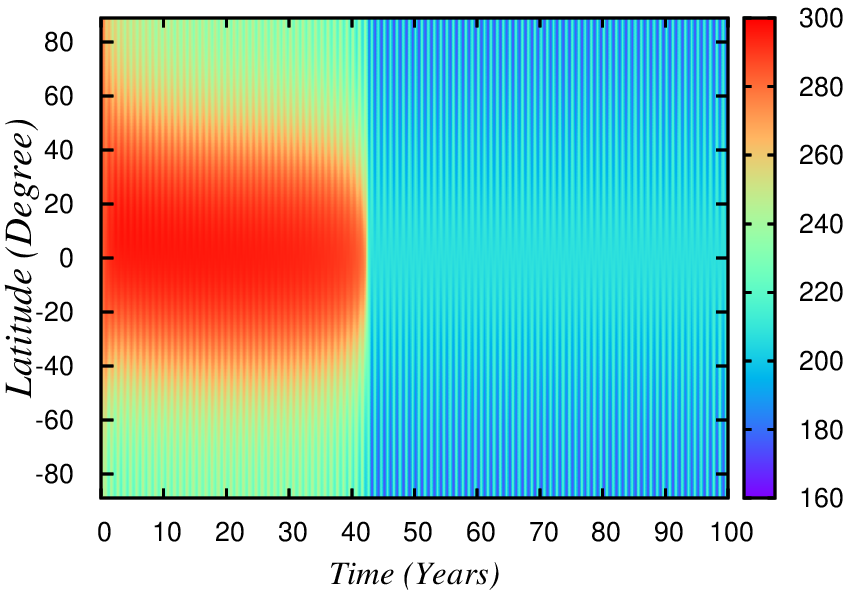}
    \end{subfigure} \\
    \begin{subfigure}{0.49\textwidth}
        \includegraphics[width=\textwidth]{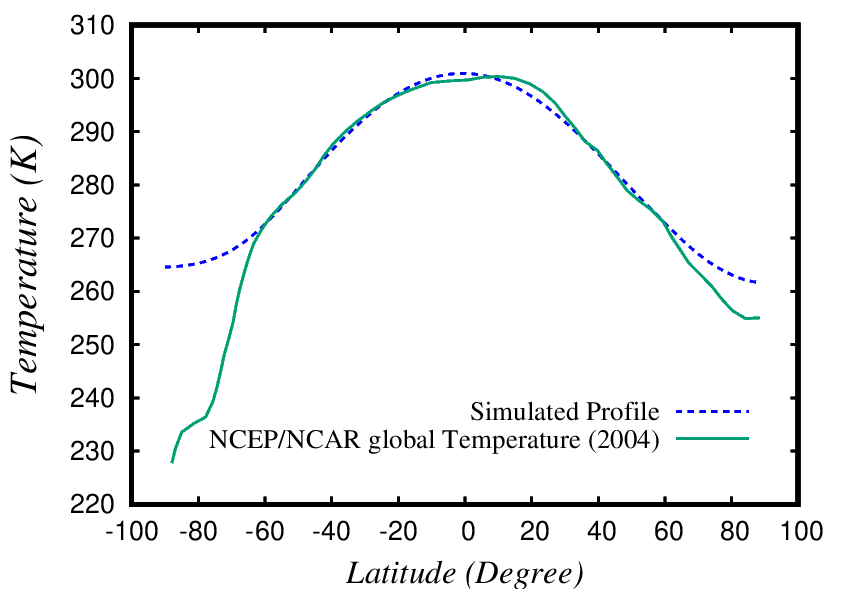}
        \caption{Temperature profile, simulation vs. data}
    \end{subfigure}
    \begin{subfigure}{0.49\textwidth}
        \includegraphics[width=\textwidth]{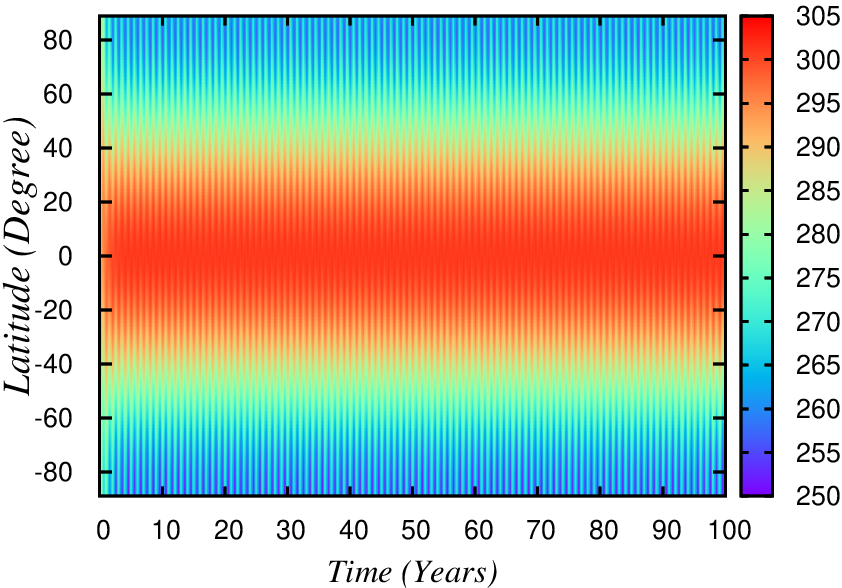}
        \caption{space time temprature profile}
    \end{subfigure}
    \newline
\caption{Dependence on initial orbital position. Top row: Global freezing occurred when the simulation started at northern solstice using albedo in Spiegel et al. (2009). Bottom row: No global freezing happened when the simulation started at northern solstice using albedo in equation \ref{r8} }
    \label{Alstart}
\end{figure}

\begin{figure}[ht]
    \centering
        \begin{subfigure}{0.49\textwidth}
        \includegraphics[width=\textwidth]{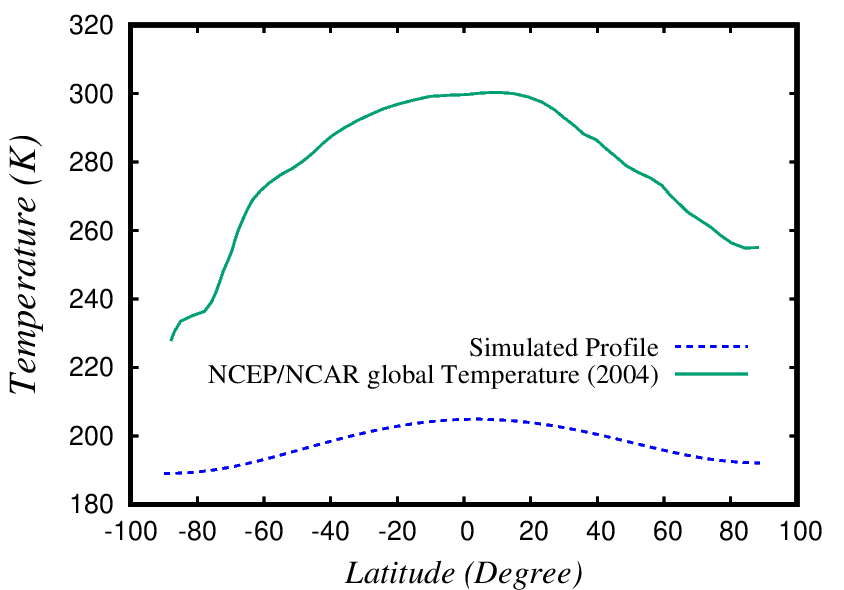}
    \end{subfigure} 
          \begin{subfigure}{0.49\textwidth}
        \includegraphics[width=\textwidth]{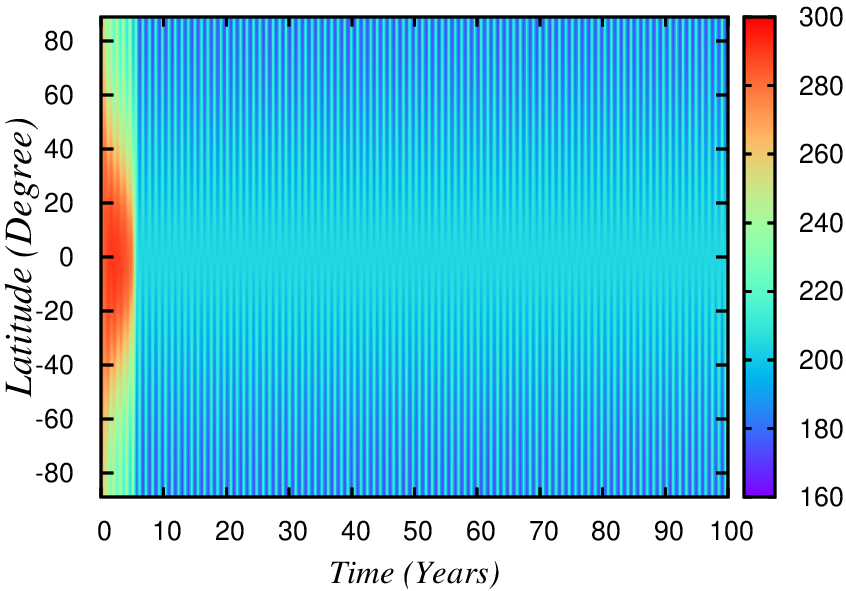}
    \end{subfigure} \\
    \begin{subfigure}{0.49\textwidth}
        \includegraphics[width=\textwidth]{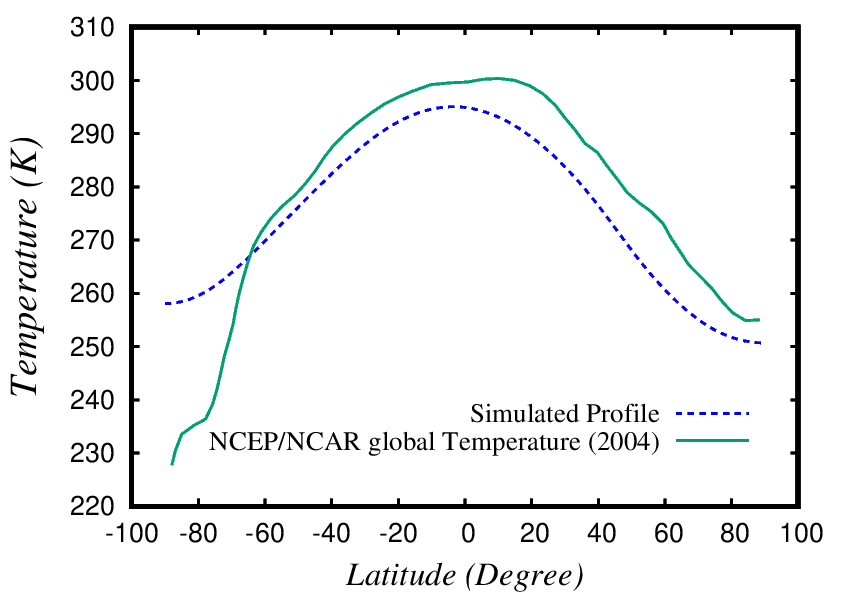}
        \caption{temperature profile, simulation vs. data}
    \end{subfigure}
    \begin{subfigure}{0.49\textwidth}
        \includegraphics[width=\textwidth]{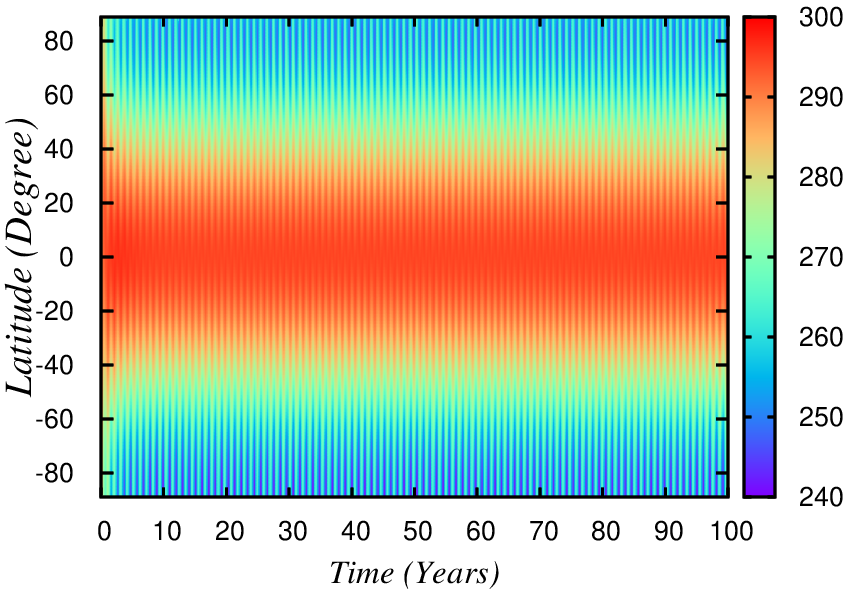}
        \caption{space time temperature profile}
    \end{subfigure}
    \newline
    \caption{Temperature profile of the Earth when its distance to the Sun is changes from 1 A.U. to 1.025 A.U. Top row: Global freezing using albedo in Spiegel et al. (2009). Bottom row: No global freezing happened using albedo in equation \ref{r8} }
    \label{Aldistance}
\end{figure}

The LEBM implemented by Forgan (2013) and Spiegel et al (2009) works well for the Earth-Sun system. However, it is highly dependent on the starting point of the simulation and the exact 1 A.U. distance of the Sun-Earth system. The problem arises from the fact that their choice for the Albedo is based on only temperature. Equation \ref{appa1} is implemented by both Forgan (2013) and Spiegel et al (2009).
\begin{equation}
A(T)=0.525-0.245 \tanh \left[\frac{T-268 \mathrm{K}}{5 \mathrm{K}}\right]
 \label{appa1}
\end{equation}
Figure \ref{Alstart} shows that by implementing \ref{appa1}, if the simulation starts at northern solstice, the planet will freeze totally after about 40 years. Also, Figure \ref{Aldistance} shows that if the Earth-Sun distance increases by only 0.025 AU, the planet will freeze after about 5 years. However, if we use the relation implemented by Vladilo et. al (2013), which is described in the Albedo subsection \ref{Albedo}, both problems in turning the planet in an Earth-Sun like system to a frozen one would be solved. The bottom rows of figures \ref{Alstart} and \ref{Aldistance} show this remedy.

\newpage

\section{Model validation}
\label{sectionthree}

\begin{figure}[H]
    \centering
    \begin{subfigure}{0.49\textwidth}
           \includegraphics[width=\textwidth]{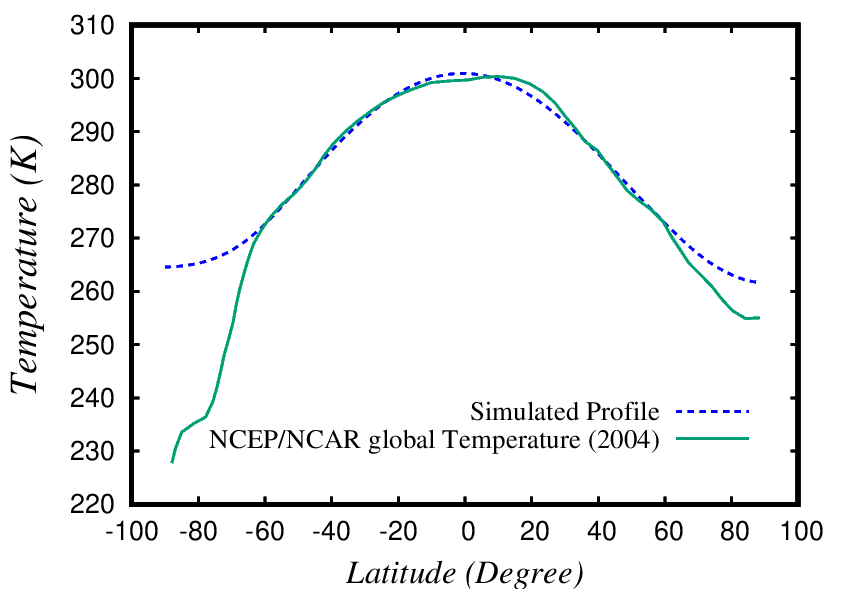}
        \caption{}
        \label{fig1}
    \end{subfigure}
      \begin{subfigure}{0.49\textwidth}
               \includegraphics[width=\textwidth]{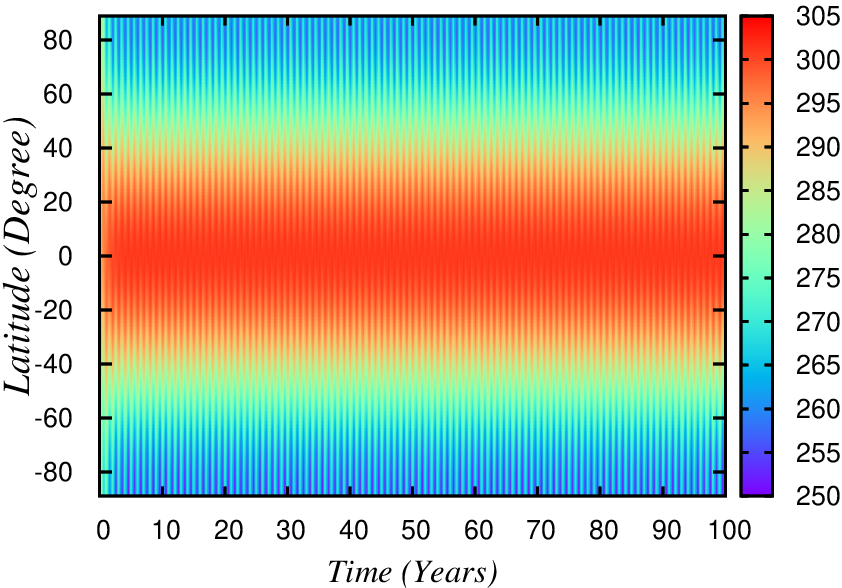}
        \caption{}
        \label{fig2}
    \end{subfigure} 
    \caption{(a) Comparing real mean data of the temperature in different latitudes of the Earth with the simulation data of a planet in an Earth/Sun-like system. (b) Temperature profile of simulated data for different latitudes for a one hundred year simulation}
    \end{figure}

We should assess the validity of the model by applying it to an Earth/Sun-like system, comparing its result with real Earth data. In figure \ref{fig1}, the green line shows the average taken from the NCEP/NCAR global temperature data and the dashed line is our model simulation for an Earth-like planet orbiting a Sun-like star. It is a good match except in the south polar region (North \& Coakley, 1979). Figure \ref{fig2} shows the temperature profile in different latitudes for a one hundred year simulation run. The decreasing temperature from the equator to the poles and the effect of seasonal changes are visible in these figures.
\nolinenumbers

\end{doublespace}
\end{document}